# Annealing of swift heavy ion tracks in amorphous silicon dioxide

Shankar Dutt[1, *], Christian Notthoff[1], Xue Wang[1,2], Christina Trautmann[3,4], Pablo Mota-Santiago[5], and Patrick Kluth[1]

[1]Department of Materials Physics, Research School of Physics, Australian National University, Canberra ACT 2601, Australia

[2]State Key Laboratory of Nuclear Physics and Technology, Peking University, Beijing 100871, China

[3]GSI Helmholtzzentrum für Schwerionenforschung, Planckstr. 1, 64291 Darmstadt, Germany

[4]Technische Universität Darmstadt, 64289 Darmstadt, Germany

[5]ANSTO-Australian Synchrotron, Melbourne, Australia

*Email: shankar.dutt@anu.edu.au

# Abstract

The annealing kinetics of the high energy ion damage in amorphous silicon dioxide (a-$SiO_2$) are still not well understood, despite the material's widespread application in material science, physics, geology, and biology. This study investigates how annealing temperature, duration, and ambient environment affect the recovery of irradiation damage produced along the trajectory of swift heavy ions in a-$SiO_2$. The track-annealing kinetics and the changing ion track morphology were investigated using synchrotron-based small-angle X-ray scattering (SAXS) and etching methods. We found that track annealing proceeds quicker near the sample surface demonstrated by a changing track etch rate as a function of depth. Measurements of ion tracks using SAXS show only small changes in the radial density distribution profile of the ion tracks. Activation energy of the annealing process at different sample depths was determined and the effect of the capping layer during the annealing process was also studied. Combination of oxygen diffusion and stress relaxation may contribute to the observed behaviour of preferential and anisotropic healing of the ion track. The results add to the fundamental understanding of ion track damage recovery and may have direct implications for materials for radioactive waste storage and solid state nanopores.

# 1. Introduction

When heavy ions with energies exceeding a few tens of MeV, also termed 'swift heavy ions', penetrate a solid, they lose their energy mainly through inelastic interactions with the electrons in the solid. In many materials, this process typically leads to the formation of narrow, straight damage trails called 'ion tracks' [1–3]. In nature, such damage trails occur as so-called 'fission tracks' in minerals that contain radioactive uranium or thorium impurities. By spontaneous fission, high energy fragments of ~100 MeV are produced, which create ion tracks, e.g., in quartz, mica, apatite or zircon [4]. Due to the damage, ion- and fission tracks are often more susceptible to chemical etchants than the undamaged surrounding matrix material. If the etching process along the ion track is much faster than the etch rate of the undamaged bulk material, high-aspect-ratio etch pits up to quasi cylindrical channels can be fabricated. The specific pore shape and size depend on a variety of factors including the ion energy, etchant characteristics, and the solid itself. The discovery of ion tracks and the fact that the tracks are more susceptible to chemical etchants as compared to the bulk matrix has sparked significant interest in many research fields including physical chemistry and material science [5–11], archaeology and geology [12–16], space science [17], micro- and nano-electronics [18–20], optics, including plasmonics [21,22], nuclear science [14], and more recently biophysics [7,23].

Typically, ion tracks are a few nanometres in diameter and depending on the kinetic energy of the incoming ions, can be up to a few tens of micrometres in length. When ion tracks are annealed, they usually shrink in size, and with increasing temperature eventually disappear completely. A detailed understanding of the annealing kinetics, structural changes, and influence of the surface on track annealing is still lacking. Along with the fundamental understanding of the ion track damage recovery, laboratory-based annealing studies are also important to validate models for fission track thermochronology. To date, annealing kinetics of ion tracks have predominantly been studied in minerals and crystalline solids [12,24–32].

Whereas most of the earlier studies used fission fragments, in recent studies ion tracks formed using heavy ion accelerators were used as they enable irradiation with well-defined energies and provide a good proxy for fission tracks [33,34]. Earlier annealing studies used ion track etching and were primarily focused on getting the correct interpretation and record of fission track dating, cosmic rays in meteoritic and lunar samples, and less attention was given to the changing ion track morphology itself.

In amorphous silicon dioxide (a-$SiO_2$), ion tracks were found to have a cylindrical core-shell structure with an under-dense core enveloped by an over-dense shell with a smooth transition between the two regions [35–37]. There have been only few annealing studies of ion tracks in a-$SiO_2$ [38,39]. These studies, however, do not reflect the exact picture of the annealing mechanism. Due to the use of fission fragments from radioactive isotopes for the fabrication of the ion tracks, these studies were also susceptible to more statistical errors inherent with fission tracks. Unlike A. Aframian [38], who annealed the a-$SiO_2$ samples to 1000 ºC and discovered no tracks for samples annealed at temperatures higher than 650 ºC, S.L. Koul *et al.* [39] have only annealed samples to 390 ºC. Although both investigations discovered that the observed track density (i.e., the number of tracks per unit area) decreases with increasing annealing temperature, there is a significant gap in knowledge about the structural changes, thermal recovery of the damage, and the variation of the ion track profile with depth. These studies also do not explain why the track density decreases. The present manuscript aims to provide significant insights into how the annealing behaviour of ion tracks in a-$SiO_2$ depends on the temperature and time, and how the annealing proceeds from the sample surface. The results from the present study should also improve our general understanding of track annealing.

This study is also of potential interest for researchers working on nuclear waste storage. More than a quarter-million metric tons of highly radioactive waste sits in storage worldwide.

Mostly, the liquid waste is vitrified in borosilicate glass and stored in steel tanks. Recently, the U.S. Department of Energy has started building a waste treatment and immobilization plant at the Hanford site using Joule-heated ceramic melters to process the nuclear waste [40]. Although the vitrification process of nuclear waste is well established, the long-term performance of the vitrified waste forms is not completely understood. The damage to the borosilicate glasses is generally caused by fission fragments, alpha-particles, and electrons affecting the chemistry, hardness as well as structure of the glass [40–43]. The methods discussed in this manuscript can also be used to investigate track damage in borosilicate glasses. Although the temperature of nuclear waste during storage depends on the activity of the waste, burial location and depth, our results could be useful for the researchers working in this field. Our results demonstrate the potential for annealing ion-tracks at relatively lower temperatures. This is particularly relevant when the waste is stored on land and easily accessible. Presently, the most commonly used method for the disposal of nuclear waste is deep geological burial. If the deep rock melting process [44] is utilized, very high temperatures, up to 1200 °C, are achieved, which are sufficient for annealing the damage completely.

Track etched membranes are of interest for filtration processes with small pore size distribution such as water-desalination, ion-pumps and many more applications. Membranes with single pores have attracted the attention of researchers all over the world for bio- and chemical-sensing and DNA sequencing [7,45–49]. Our results show that through the annealing of ion tracks, the shape of etch pits in $SiO_2$ can be adjusted. Hence, this study also presents a new approach for tailoring the shape of solid-state nanopores for such applications.

To study the influence of annealing on the ion tracks, we utilized chemical etching complemented by synchrotron based small-angle X-ray scattering (SAXS). Chemical etching was used to convert the ion tracks into etch pits to learn about the profile of the ion tracks and their variation with depth and temperature. As the etching of ion tracks depends directly on the track structure, this method provides first-hand information about the ion tracks [50,51]. SAXS on the other hand measures the radial density profile of the annealed ion tracks directly. SAXS is a non-destructive powerful technique based on the elastic scattering of monochromatic X-rays from density fluctuations within a solid at a nanometre scale [52–54]. SAXS allows accurate and statistically reliable information about the shape and size of the ion tracks, particularly for parallel and monodisperse systems [13,35,36,50,55]. In the present manuscript, we have studied different aspects of the annealing of ion tracks in a-SiO$_2$ by varying annealing temperature, duration as well as ambient atmosphere. The experimental results have been supported by calculations of the etch pit shape that enable a discussion of how the ion track profile changes with depth.

## 2. Experimental Procedures

*2.1 Ion irradiation*

The ion tracks were produced in 2 µm thick a-SiO$_2$ grown on Si (100) by wet thermal oxidation of silicon (acquired from WaferPro LLC, US) [56] by irradiation with 1.6 GeV Au ions at the UNILAC accelerator facility at the GSI Helmholtz Centre for Heavy Ion Research, Darmstadt, Germany. Fluences ranged between $1\times10^8$ ions/cm$^2$ and $5\times10^{11}$ ions/cm$^2$. The irradiations were performed at room temperature and the direction of the incoming incident ions was normal to the surface of the sample. The maximum flux density during irradiation was 2 x 10$^8$ ions cm$^{-2}$ s$^{-1}$. Under these conditions sample heating during irradiation is considered negligible. The average electronic energy loss $S_e$ in the SiO$_2$ layer calculated using the SRIM2008 code [57] is 21.1 keV/nm and remains almost constant throughout the film ($\Delta S_e \leq 0.4\%$) (see figure S1). The ion tracks are thus considered homogenous throughout the sample thickness, i.e., with no significant variations along their axes.

*2.2 Ion track annealing*

The irradiated samples were annealed in a temperature-stabilized electric tube furnace at different temperatures and for different durations. The temperature was allowed to stabilize for a minimum of one hour before the samples were inserted into the furnace. Three different strategies were employed:

   a. Irradiated a-$SiO_2$ samples were annealed under nitrogen atmosphere at temperatures ranging from 200 ºC to 900 ºC in increments of 50 ºC for 30 min. These experiments provide insights into the dependence of ion track annealing as a function of temperature.

   b. Irradiated a-$SiO_2$ samples were annealed for various periods of time while maintaining a constant annealing temperature. The annealing time was varied from ~1 min to 3000 min to study the influence of annealing time on track recovery.

   c. Irradiated a-$SiO_2$ samples were annealed under three distinct ambient conditions, nitrogen, oxygen, and forming gas (a mixture of 95% nitrogen and 5% hydrogen gas), respectively, for a duration of 30 min at 500 °C. This strategy was used to study the impact of the annealing atmosphere (neutral, oxidizing, and reducing) on the ion track recovery. For additional samples, a ~150 nm thick aluminium metal layer deposited on the top of the samples served as a capping layer. Irradiated samples with and without capping layer were then annealed at 500 °C for 30 mins to study the impact of the oxygen on the annealing process of ion tracks.

*2.3 Ion Track Etching*

Samples irradiated with low fluences ($1 \times 10^8$ and $5 \times 10^8$ ions/cm$^2$) were etched at room temperature in 3% hydrofluoric acid (HF) for 10 min. Samples were kept in a vertical position using a custom-built etching cradle to warrant reproducible etching conditions. To stop the etching process, samples were removed from the etching bath, rinsed multiple times in fresh deionized water and subsequently dried in air. The etching process leads to the formation of conical etch pits in the samples [51]. The etching threshold value of electronic energy loss for continuous and uniform formation of etch pits in a-SiO$_2$ has been reported to be ~4 keV/nm [58]. As detailed in section 2.1, under the present irradiation conditions, the electronic energy loss is calculated to be 21.1 keV/nm, a value significantly higher than the reported threshold.

Plan-view and cross-section scanning electron microscopy imaging of the etch pits was performed using a FESEM instrument (ThermoFisher - FEI Verios 460L). Low fluence ($1\times10^8$ and $5\times10^8$ ions cm$^{-2}$) irradiated samples were used for this process to minimize the overlap of the etch pits. Radii were deduced from plan-view SEM images measuring on average ~200 etch pits to obtain good statistics.

We measured the bulk etch rate of different annealed and unannealed samples using JA Woollam M-200D Ellipsometer and Bruker Dektak® surface profiler. By means of ellipsometry, the decrease in thickness of the a-SiO$_2$ film after a certain etching time was calculated and for surface profiler measurements, an etch stop medium (Apirzon wax and polyimide film) was used to cover part of the sample and height difference between the etched and unetched part was measured. Independent of the annealing conditions, and using both measurement techniques, the bulk etch rate was found to be $15.6 \pm 0.6$ nm min$^{-1}$. The track etching rate was determined through the etching model (see section 2.5) using the bulk etch rate and the etch pit radius. Other methods to measure the track etch rate include conductometry during the etching process [59] and through the analysis of three-dimensional wall profiles of the etch pits [60]. Conductometry is only possible for membranes and require the substrate to be removed.

*2.4 Reactive Ion Etching*

To study track annealing as a function of depth, the annealed and unannealed a-SiO$_2$ samples were etched from the surface using a Samco 400iP (Japan) Inductively Coupled Plasma-Reactive Ion Etching (ICP-RIE) instrument. The etching process was carried out at a flow rate of 10 sccm of CHF$_3$. Part of each sample was protected during the etching process by covering it with a polyimide film. By surface profilometry, the etch rate of the ICP-RIE process was determined to be 53.7 $\pm$ 3.3 nm min$^{-1}$, regardless of the annealing conditions.

To exclude a possible impact of reactive ion etching on the wet etching of the ion tracks, untreated and ICP-RIE treated unannealed irradiated samples were etched for 10 minutes in a 3% HF solution. Comparing the etch pits in both samples, we found no size and shape difference and thus conclude that ICP-RIE has no effect on the ion tracks.

*2.5 Track etching model*

Understanding the ion track etching process and which parameters impact the geometry of the etch pits is necessary for evaluating the results from the etching experiments. Figure 1 shows a schematic of the geometry and parameters involved in etch pit formation. The track etching process, i.e., the conversion of an ion track into an etch pit is predominantly governed by the two competing etching rates, (i) the track etch rate ($V_T$), which describes the dissolution of the damaged material along the ion track and (ii) the bulk etch rate ($V_B$), which describes the rate at which the undamaged bulk matrix is etched isotropically. The shape of the etch pit is determined by the ratio of $V_T$ to $V_B$ and pit formation is not possible if $V_B \geq V_T$. As illustrated in Figure 1, the half cone angle $\delta$ of the etch pit depends on $V_B$ and $V_T$ only. $D$ is the initial thickness of the thin film and after etching for time $t$, it reduces to $D - h$ where $h = L' - L$ is the thickness of the bulk layer removed during the etching process ($h = V_B t$). From Figure 1 we see:

$$\sin \delta = \frac{V_B}{V_T} \qquad \text{Eq. 1}$$

and,

$$\tan \delta = \frac{V_{R_1}}{V_T - V_B} = \frac{V_R - r/t}{V_T - h/t} \qquad \text{Eq. 2}$$

where $V_R = R/t$ is the radial etch rate, i.e., the etch rate parallel to the surface. We assume that the ion track with radius r etches homogeneously with the track etch rate. It is worth noting that the radial etch rate $V_R$ is not an independent variable but is derived from $V_T$ and $V_B$ through the geometry of the etch pit, as the etch front moves perpendicular to the cone surface. This means that the etch pit openings at the surface grow slower for slower track etch rates. This relation is clear from Eq. 2 and can be written as:

$$V_R - \frac{r}{t} = \tan \delta \cdot (V_T - V_B) \qquad \text{Eq. 3}$$

Combining Eq. 1 and Eq. 3, we can determine the track etch rate as a function of the bulk etch rate $V_B$, etch pit radius R, and ion track radius r:

$$V_R - \frac{r}{t} = \tan\left(\sin^{-1}\left(\frac{V_B}{V_T}\right)\right) \cdot (V_T - V_B)$$

$$V_R - \frac{r}{t} = \frac{V_B \cdot (V_T - V_B)}{\sqrt{V_T^2 - V_B^2}} \qquad \text{Eq. 4}$$

$$\Rightarrow V_T = \frac{V_B(t^2 V_B^2 + (r - tV_R)^2)}{(t(V_B - V_R) + r)(t(V_B + V_R) - r)}$$

Using $V_R = R/t$ and simplifying above equation, we get:

$$V_T = \frac{V_B(t^2V_B^2 + (R-r)^2)}{t^2V_B^2 - (R-r)^2} \qquad Eq.\ 5$$

Using Eq. 5, we can determine the track etch rate as a function of the bulk etch rate, etch pit radius R, and ion track radius r. These parameters are all accessible through the experiment. The model derived here directly applies to a constant track etch rate leading to the formation of a conical etch pit. For the case of a variable track etch rate, $V_T$ can be expressed by $dV_T$ and $\delta$ by $d\delta$ and we calculate the parameters at intervals of 1 nm. In this case, the surface of the etch pit will be curved.

## 2.6 Activation Energy of track annealing

Many studies in the past have investigated the annealing kinetics of tracks in different crystalline solids, minerals, and polymers [29,61–64]. To characterize the annealing behavior of ion tracks in a-SiO2 at different depths, we employed a semi-empirical relationship reported by P.B. Price *et al.* based on their results of thermal annealing in phosphate glasses [61]:

$$\frac{(S-1)_i - (S-1)_f}{(S-1)_i} = At_a^{1-n}\exp\left(\frac{-E_a}{k_B T}\right) \qquad Eq.\ 6$$

where $S$ is the ratio of the track and bulk etch rate, $S = V_T/V_B$ and the subscripts $i$ and $f$ correspond to the unannealed and annealed values respectively. $t_a$ (min) and $T$ (K) are the annealing time and annealing temperature, respectively; $k_B$ (eV/K) is the Boltzmann constant, $A$ is a proportionality constant and $E_a$ (eV) is the activation energy for the annealing process. For our analysis we chose a constant annealing time ($t_a = 30$ min) for all samples, such that Eq. 6 can be simplified to:

$$\ln\left(\frac{(S-1)_i - (S-1)_f}{(S-1)_i}\right) = -\frac{E_a}{k_B T_a} + A' \qquad Eq.\ 7$$

$A'$ in the above equation includes all constants. By plotting $\ln\left(\frac{(S-1)_i-(S-1)_f}{(S-1)_i}\right)$ vs. $\frac{1}{k_B T_a}$, and fitting the data, the activation energy of the annealing process can be deduced from the slope of the curve.

## *2.7 Small angle X-ray scattering*

The density profile of un-etched ion tracks was studied using synchrotron-based SAXS [13,32,35,36]. Transmission SAXS was performed at the SAXS/WAXS beamline at the ANSTO Australian Synchrotron. Samples irradiated with fluences of $5 \times 10^{11}$ ions/cm² were employed for the SAXS measurements in order to achieve sufficiently high scattering intensities and good signal to noise ratios. At this fluence, track overlapping effects are negligible as calculated using a stochastic overlap model [65,66]. X-rays of energy 12 keV (wavelength = 1.03 Å) were used and sample to detector distance was 1607 mm. To enable precise alignment and defined rotation with respect to the incoming X-rays, the samples were mounted on a three-axis goniometer. Firstly, the ion tracks were approximately aligned with the incoming X-ray beam. The corresponding scattering pattern shows symmetric circular scattering pattern as shown in Figure 2 (a). Subsequently, samples were titled by an angle of 5° or 10° exhibiting a highly anisotropic pattern resulting from the high aspect ratio of the aligned ion tracks (Figure 2 (b)). The straight bright lines from the crystalline Si substrate are ascribed to Kossel lines [67] and were masked during the data extraction process. For analysis, the two-dimensional scattering patterns were reduced into a one-dimensional intensity pattern. This was done by performing an azimuthal integration along the streak applying a narrow mask followed by background subtraction using an area away from the streak [35,36,55]. Following data reduction, a core-shell model [36] was used to fit the one-dimensional scattering data to unearth the information of the average ion track density profile in un-annealed and annealed samples. Using a non-linear least-square fitting algorithm [68], the fitting of the 1D SAXS data obtained after data reduction was performed using a Python and C-based custom code.

# 3. Results

## 3.1 *Ion track annealing near the sample surface*

To study how the track morphology changes with annealing temperature, we etched the irradiated samples (fluence = $1 \times 10^8$ ions/cm$^2$) using the procedure described in the previous section. The resulting conical etch pits were characterized using plan-view as well as cross-section SEM imaging. It should be noted that during the etching process, the top ~156 nm of the a-SiO$_2$ thin film were removed. The results thus present an average over the top ~156 nm plus the depth of the etch pit. For this reason, we relate the results to changes near the sample surface.

Figure 3 shows plan-view and cross-section SEM images of the etch pits, together with the surface radii (obtained from plan-view images, represented by solid blue squares) and half cone angle values (obtained from the cross-section SEM images, represented by solid red circles). The SEM images are taken at the same magnification to better visualise the changing etch pit size and shape. The solid lines are a guide to the eye. The radius of the etch pits and their depth decrease with increasing annealing temperature, while the measured half cone angle increases. For samples annealed at 200 °C, etch pit dimensions and morphology are the same as those of the unannealed sample. In contrast, samples annealed within the 600 °C to 900 °C range exhibit no discernible etch pits, suggesting that either their size falls below the detection threshold of scanning electron microscopy, or the ion track damage has undergone adequate restoration, thus preventing the formation of etch pits.

As previously stated, the ratio of the track to bulk etch rate determines the shape and size of the etch pit. Given the bulk etch rate remains constant upon annealing, our findings suggest that the track etch rate and thus the structure of the ion tracks changes upon annealing. Interestingly, the trend of decreasing etch pit radius is quite similar to the decrease in track retention efficiency with increasing annealing temperature reported by A. Aframian [38]. In contrast, however, we do not observe any decrease in the number density of the etch pits in our a-$SiO_2$ samples. We counted the number of etch pits in an area of 21 $\mu m$ x 14 $\mu m$ for all the annealed and unannealed samples and in effect found no difference in the number density of the etch pits. The measured number density of the etch pits also directly correlates with the ion fluence. This disparity can possibly be explained by the statistical variation in energy and direction inherent to the fission tracks and the method of data evaluation used by A. Aframian.

## 3.2  *Ion track annealing as a function of depth*

Eq. 5 was utilized to calculate the track etch rate from the radii of the etch pits determined by scanning electron microscopy measurements, assuming a constant track etch rate. The radius of the ion tracks was determined using SAXS (explained in section 3.5) to be $4.6 \pm 0.1$ nm for these calculations. After calculating the track etch rate, we compared the depth of the etch pits observed through cross-section SEM with the calculated values.

The calculations for an unannealed sample yield a track etch rate of $47.5 \pm 4.1$ nm min$^{-1}$. Considering the bulk etch rate and using Eq. 5, the depth of the etch pit for the unannealed sample was calculated to be $317.2 \pm 34.7$ nm. This value agrees well with the experimentally observed mean depth of $321.4 \pm 5.2$ nm, measured from the cross-section SEM images (cf. Figure 3). The same calculations for the sample annealed at 500 °C for 30 min yield a track etch rate of $18.7 \pm 0.8$ nm min$^{-1}$, resulting in a calculated depth value of $28.8 \pm 8.9$ nm. However, this computed value differs significantly from the experimentally measured value of 73.2 ± 6.7 nm. This discrepancy between the calculated and experimental cone depths occurs for all annealed samples and increases with increasing temperature. This deviation is apparent from Figure 3, which shows the calculated values of the half cone angle (represented by open red circles) and the measured values (represented by the solid red circles). As the calculations assumed a constant track etch rate, the deviations indicate that the track etch rate, and hence the ion track profile, varies with depth and is temperature dependant after annealing above 200 °C.

To assess how the track annealing changes with depth, we annealed an a-SiO$_2$ sample irradiated with 1.6 GeV Au ions with a fluence of $1 \times 10^8$ ions cm$^{-2}$ at 500 °C for 30 min in N$_2$ atmosphere. After annealing, half of the sample was covered with a polyimide film and etched using the ICP-RIE procedure for a duration of 6 mins. The polyimide film protects the a-SiO$_2$ layer underneath from etching and ~330 nm of the unprotected layer was removed during this process. Subsequently, the protective film was removed, and the sample was analysed by profilometer scans from the unprotected to the protected borderline. Figure 4 (a) shows such a scan with a clear step of ~330 nm in height. The sample was then etched for 10 min in 3% HF to convert the ion-tracks into etch pits. Afterwards, we analysed the size of etch pits using SEM and found a significant difference (72.5% relative change in radius) between the ICP-RIE etched and unetched sides. Figure 4 (b) shows a plan-view SEM image that extends over both sides. It is evident from the image, that on the ICP-RIE etched side, the etch pits are much larger (radius = 93.5 ± 2.7 nm) compared to those on the unetched side (radius = 54.2 ± 3.2 nm). These results are a strong indication that the ion track etch rate changes with depth. Although the size of the etch pits increases with depth, it does not reach that of the unannealed sample. We processed another sample annealed at 500 °C, where we removed ~590 nm of the top layer using ICP-RIE and observed a similar radius to the sample where ~330 nm was removed. This seems to indicate that the track etch rate levels out at certain depths.

Our findings indicate that the effect of annealing on the ion track etch rate (and hence on its structure) proceeds quicker near the surface until it saturates at a certain depth. This bodes the question if this effect also exists at the interface between the a-SiO$_2$ layer and the silicon substrate. To determine this, we removed ~1550 nm and ~1690 nm of the sample from the top using ICP-RIE before etching with HF. Considering the bulk etch rate, we are only ~300 nm and ~150 nm away from the SiO$_2$/Si interface. We discovered that the radius is comparable to that seen in samples with the top layers of thickness ~330 nm and ~590 nm removed, indicating that the ion track annealing is not enhanced near the SiO$_2$/Si interface.

A schematic of the changing ion track annealing profile is illustrated in Figure 5 (a) for unannealed ion tracks and ion tracks annealed at 300 °C and 500 °C. As discussed earlier, and as illustrated in Figure 5 (a), the ion track profile changes with depth up to a certain point. The depths at which the change saturates is dependent on the annealing time and temperature. To investigate the rate of change of the ion track annealing profile with depth, and the effect of annealing temperature and time, we performed a systematic series of experiments as illustrated in Figure 5 (b). Irradiated a-SiO$_2$ samples were annealed at different temperatures (300 °C, 350 °C, 500 °C, and 550 °C) for different durations. The annealing times were 10 min, 30 min, and 90 min for all temperatures except for annealing at 500 °C where we annealed the samples for 1 min, 10 min, 30 min, 90 min, 360 min, and 3000 min. Following annealing, surface layers were removed in nine steps of 53.7 ± 3.3 nm using the ICP-RIE process. This procedure allowed us to evaluate the ion track profile at 10 different depths. The ICP-RIE processed samples were then etched in 3% HF for 10 min. As a reference, an unannealed sample was processed in the same way. Figure 5 (c) shows a photo of an ICP-RIE processed a-SiO$_2$ sample. The different layer colours observed in the photo correspond to different a-SiO$_2$ thicknesses after ICP-RIE etching.

The radius of the etch pits measured by SEM as a function of initial depth is shown in Figure 6. As a reference, all graphs show the radii of the etch pits for the unannealed sample. It is noteworthy that in this and the next sections, we use the term "initial depth" to describe the depth of the ICP RIE processed sample before it was etched with HF. As mentioned in the experimental section, HF etching causes bulk etching of the material, and the final depth at which etch pits can be seen is equal to the sum of the initial depth and thickness of the bulk layer etched by HF which is 156 ± 6 nm.

The etch pit radius at every depth for the unannealed sample is almost the same indicating that the ion tracks in the unannealed sample do not vary with depth. Plots corresponding to the samples annealed at a temperature of 300 °C and 350 °C are shown in Figure 6 (a). The radii increase with increasing depth and become almost constant at a depth of ~200 nm and ~250 nm for samples annealed at 300 °C and 350 °C, respectively. For longer annealing times, the radii of the etch pits decrease in a non-linear manner. At higher depths, the difference between 30 min and 90 min annealing becomes very small. Figure 6 (b) presents the radius of the etch pits for the samples annealed at 500 °C. In this case, we have annealed samples for an extended time, ranging between 1 min and 3000 min to observe the initial changes in the damage recovery of the ion tracks as well as the annealing at very long time. We observed for the case of the sample annealed for 1 min at 500 °C that the radius changes up to an initial depth of ~150 nm and deeper in the sample, the radii are close to those of the unannealed sample. The ion track healing process proceeds very quick at the surface of the sample. For longer annealing durations (10 min, 30 min, 90 min, and 360 min), a similar trend as at 300 °C and 350 °C is observed except that the initial depth at which the etch pit radius nearly saturates is ~400 nm. We also annealed a sample at 500 °C for a duration of 3000 min. For this sample, etch pits were only observed deeper than initial depth of ~100 nm. At shallower depths, either the etch pits were too small to be observed or the ion tracks have recovered near the surface such that no etch pits were generated. This effect may lead to the erroneous conclusions that the ion tracks are annealed in the entire sample and may have been misinterpreted in previous studies [38,39].

Figure 6 (c) shows the evolution of the radius with depth for samples annealed at 550 °C between 10 and 90 min. Similar to samples annealed at lower temperatures, we observe an increasing radius with depth, and we do not observe any etch pits for depths less than ~150 nm for the case of the sample annealed for 90 min. To compare the influence of temperature directly, we plot the radius of the etch pits as a function of depth for different temperatures for an annealing time of 30 min in Figure 6 (d). The data clearly show that the annealing time and temperature directly influence the depth at which the radius variation becomes very small as well as track damage recovery and thus the etching rate.

To obtain quantitative information about the track-healing process, we fit the radius as a function of initial depth using the Gompertz function [69]:

$$R(d) = A \cdot \exp(-\exp(-k \cdot (d-c))) \qquad \text{Eq. 8}$$

where, d is the initial depth, A is the amplitude of the curve, c and k define the center and the slope of the curve. The fits to the data are shown in Figure 6 as solid black line and the fit parameters obtained are listed in table S1. From the parameters obtained, Eq. 8 yields the radius as a function of depth for each annealing temperature and time. We can also deduce the radial etch rate as a function of depth: $V_r(d) = R(d)/t$ where t is the etching time (t = 10 min for this data) and then use the values for the radial etch rate in Eq. 4, and obtain the track-etch rate as a function of initial depth (figure S2). As expected, the trend of the track-etch rate as a function of annealing time and temperature is identical to that of the radius with initial depth as these parameters are directly correlated.

Figure 7 shows three-dimensional surface plots depicting the track etch rate as a function of annealing time and initial depth of the sample for samples annealed at 300 °C (a), 350 °C (b), 500°C (c), and 550 °C (d). We have considered the track etch rate of an unannealed sample for the annealing time (t) = 0. The track etch rate initially decreases quickly and with increasing annealing time the change becomes smaller. It is also apparent that the track etch rate increases with increasing initial depth of the sample and seems to level off at a particular depth depending on the temperature and annealing time. Further examination reveals that at the surface, the change in track etch rate, and hence the change in ion track morphology, is greater than at any depth. The rate of change falls dramatically in the first few minutes, indicating that the majority of track healing occurs within the first few minutes of the annealing process and that the ion track annealing proceeds significantly slower after that. A comparison of the track etch rate as a function of annealing time when the initial etched depth of the sample is 0 nm and ~500 nm is shown in figure S3 (a) and (b), respectively, for different temperatures. This is essentially a cut through the graphs in Figure 7 to illustrate the exponential decrease and saturation with increasing time. The fit parameters resulting from the fit to the data using an exponential decay function are given in table S2. For all annealing conditions, the track etch rate declines more rapidly near the sample surface than at a depth of around 500 nm.

Using the variable track etch rates obtained from the fitting parameters of equation 8, we can calculate the shape of the etch pits and compare them with the results of the experiment. The simulated etch pit shape for different annealing temperatures (unannealed, 300 °C, 350 °C, 500 °C, and 550 °C) at a fixed annealing time of 30 min is shown in Figure 8 (a). These calculations correspond to the etch pits which we would get after etching the annealed and unannealed samples for 10 min in 3% HF. The etch pits start at -156 nm because of the removal of the top layer by bulk etching. Extrapolated values of the track etch rate from the analytical fits were used for simulating the etching from the surface. The simulations clearly indicate that the shape of the etch pits becomes slightly curved because of changing track-etch rate with depth. Because of the resolution limitations and charging of the sample, it is difficult to observe this curvature in cross-section SEM images. Also, the greatest curvature is observed at the top of the surface as the rate of change of ion track profile is highest near the surface. However, due to the removal of bulk material at the surface during etching, this curvature is mostly gone.

To quantify the agreement of the etching model and the experiment, we calculated the half cone angles from equation 2 using the variable track etch rates. Given the slightly curved walls of the etch pit, the angle is taken as the average over the curvature. As discussed earlier, when we calculated the half cone angle values considering the track etch rate to be constant, they did not match with the experimentally measured values. All half cone angle values are shown in Figure 8 (b). This includes experimentally determined values represented by solid black squares, calculated values considering the track etch rate to be constant, represented by solid red circles, and calculated values considering the track etch rate to be variable, represented by solid blue triangles. The half cone angle values calculated by using the variable track etch rate agree well with the experimental measured values and this agreement further validates our findings. Based on these results, we produced a video (V1) which shows the progression of fabrication of etch pits by etching the unannealed ion tracks and ion tracks annealed at 300 °C, 350 °C and 500 °C for a duration of 30 mins. Also, ion tracks annealed at 500 °C for 1 min and 3000 min are considered in this video. The removal of the bulk of the material is also considered and shown. This video clearly shows how various etch pits of different shapes and sizes with conical and curved conical forms are created, as well as how the curvature of the etch pits is diminished if the ion-tracks are etched for an extended period of time.

### 3.3 *Activation Energy of ion-track recovery*

The decreasing etch-pit radius and the growing half cone angle with increasing annealing temperature directly indicate that the ion track structure has changed. Also, the size of the etch pits changes with depth, providing evidence that the track recovery process changes as a function of depth. The activation energies for the recovery process at different initial depths (0 nm, ~210 nm and ~500 nm) are derived from Arrhenius plots (Figure 9) using the calculated values of the track etch rate for unannealed and annealed samples at the respective depths. From the linear fits of $\ln\left(\frac{(S-1)_i - (S-1)_f}{(S-1)_i}\right)$ versus $\frac{1}{k_B T}$, we obtain activation energies of $0.108 \pm 0.010$ eV, $0.157 \pm 0.024$ eV and $0.204 \pm 0.038$ eV respectively for the cases of initial depths of 0 nm, ~210 nm, and ~500 nm. These results show that the activation energy for healing ion tracks at the surface is only about half the value compared to deeper region. These results may also indicate that the healing process at the surface is different to that deep inside the sample. The annealing of ion tracks is a rather complex process. Although the track etch rate reduces with thermal annealing, as pointed out by S. Klaumünzer [11] the activation energy and its meaning generally depend on the experiment and the method of data evaluation. Thus, the values reported here do not explain the full kinetics of the annealing process, however, they could be useful in future research, particularly when comparing the annealing kinetics of ion tracks in other amorphous materials.

### 3.4 *Effect of annealing atmosphere*

To study the influence of annealing environment on the ion track recovery, three different atmospheres were used, neutral atmosphere (dry $N_2$ gas), reducing atmosphere (forming gas, a mixture of 95% dry $N_2$ and 5% dry $H_2$ gas), and oxidizing atmosphere (dry $O_2$ gas). The samples were annealed at 500 °C for a duration of 30 min. Subsequently, the samples were thinned down using ICP-RIE to expose different depth-profiles of the ion track and subsequently etched in HF. Figure 10 (a) shows the radius of the etch pits as a function of initial depth for the three different annealing atmospheres. The radius of the etch pits is ~8% and ~12% smaller for annealing in an oxygen atmosphere than for annealing in a nitrogen atmosphere and in forming gas atmosphere, respectively, at the initial depth of 0 nm. This difference between the radius values is observed up to a depth of ~200 nm. At depths larger than 200 nm, the radii are approximately the same for all annealing conditions. This discrepancy indicates that ion track damage anneals somewhat quicker in oxidising environments and implies that diffusion of oxygen from the atmosphere plays a crucial role in the repair of the track damage.

To determine whether the observed variation in the ion track profile with depth is primarily due to oxygen diffusion from the atmosphere or from the surface or if other processes contribute to the atomic-level repair of ion tracks, we deposited a ~150 nm thick capping layer of aluminium on top of the sample surface before annealing. To minimize the influence of atmospheric oxygen during annealing, these experiments were carried out in a dry nitrogen atmosphere. The Al film was RF sputtered using ATC 2400-V Sputter Coater system by AJA International, Inc and was intended to limit diffusion during the annealing process. After annealing the sample at 500 °C for 30 mins under a dry $N_2$ atmosphere, the capping layer was removed. A sample without a capping layer was processed simultaneously as reference. Afterwards, the samples with and without capping layer were processed using ICP-RIE and then etched and inspected using SEM. The results of this experiment are summarized in Figure 10 (b). Samples annealed with a capping layer show higher surface radii than those without. The difference is observed up to initial depths of ~350 nm. These findings provide further evidence of the influence of gas diffusion in the healing process. However, it is obvious that even for samples with a capping layer, the radii are not constant but increase with increasing depth (Figure 10 (b). These findings thus demonstrate the significance of diffusion in the annealing process and the importance of oxygen while also suggesting that diffusion from the surface or the atmosphere is not the only process contributing to the atomic level repair of the ion tracks that results in preferential annealing from the surface.

### 3.5  *Small-angle X-ray scattering from annealed ion tracks*

In this section, we will discuss the SAXS measurements of un-etched ion tracks before and after annealing at temperatures of 350 °C, 500 °C, and 600 °C. Figure 11 shows the corresponding scattering patterns observed for samples titled by either 5° or 10° with respect to the incoming X-ray beam for the case of unannealed ion tracks (a), and ion tracks annealed for a duration of 30 mins at a temperature of 350 °C (b), 500 °C (c), and 600 °C (d). The anisotropic pattern observed is due to the high aspect ratio of the ion tracks. We previously established that the ion tracks in a-$SiO_2$ consist of a cylindrical core-shell structure with a low density core and high-density shell [35]. Later we demonstrated that between the core and shell regions, there is a gradual density transition [36]. The radial density distribution of the ion tracks defined by this model (the Core Transition Shell model) is shown in Figure 11 (e). The model is characterized by under/over- or over/under- dense core region ($R_c$) and shell region ($R_s$). The smooth transition between the electron densities of both core and shell regions is given by the transition region ($T_t$). We implemented a narrow Schulz-Zimm distribution to account for the dispersity in the density distribution profile of the ion tracks [36,55].

The SAXS measurements yield the radial density profile of the ion track averaged over the entire length of the track. To study the homogeneity of the ion track radial density, we recorded SAXS patterns from samples from which top layers were removed in a controlled manner by ICP-RIE etching. We did not observe a measurable difference between the shapes of different intensity patterns. This implies either there is no significant change in the radial density with depth, or the density variation near the surface is sufficiently reduced that it has no influence on the average radial density. To differentiate between the two options, we studied the relative scattering intensities of the patterns. The percentage change of length of ion tracks and square root of corresponding observed intensity as a function of the thickness of the layer removed from the samples is shown in figure S4. We discovered that the reduction in observed SAXS intensity with removal of top layers of sample scales with the reduction in the total length of the ion tracks. This suggests that the ion track density distribution remains nearly constant throughout the sample depth. As a result, SAXS measurements and chemical etching of ion tracks of different layers of sample annealed at the same temperature show that, while ion track damage heals faster from the surface (possibly due to the restoration of different bonds), the density profile of the tracks does not vary significantly over the depth of the tracks.

For the following discussion, we are thus considering scattering patterns from complete unannealed/annealed ion tracks. The reduced scattering intensities obtained from unannealed ion tracks as well as ion tracks annealed at different temperatures, along with their numerical fits (solid red line) using the Core Transition Shell model, are shown in Figure 12. It is apparent that the model reproduces the scattering curves with high accuracy. The fitting parameters obtained from the SAXS measurements, i.e. core region ($R_c$), shell region ($R_S$), the ratio of the density of the shell region ($\rho_s$) to the density of the core region ($\rho_c$), dispersity ($\sigma$), and radius of ion tracks ($R_c + T_t + R_s$) are listed in Table 1. During the fitting process, we fit the ratio of the density of the shell region to that of the core region while considering the density of the core region to be constant to eliminate an extra fitting parameter.

Figure 13 (a) shows the variation of the core region, transition region, shell region, and total radius obtained from the SAXS fits with increasing annealing temperatures. Above 350 ⁰C, the size of the core region decreases at the expense of an increase in the size of the transition region suggesting viscous flow between the two regions. Similarly, the shell region becomes smaller at the expense of the transition region, indicating a similar flow in these regions. Interestingly, the overall radius of the ion tracks remains almost constant, however the dispersity ($\sigma$) values slightly increase for higher annealing temperatures. The ratio of the density of the shell region ($\rho_s$) to the density of the core region ($\rho_c$) as a function of annealing temperature is presented in Figure 13 (b). The negative values of the density ratio indicate and support the earlier description of an under/over dense core/shell region. Figure 13 (c) shows the radial density distribution for unannealed and annealed tracks in a-SiO$_2$ deduced from the SAXS fitting results. The changing density profile with increasing annealing temperature is a result of the material flow into the track core from the shell of the track. The change, however, is not very large, i.e., the general core-shell structure remains. This is not surprising considering the high viscosity of a-SiO$_2$ at these temperatures. We note that we have not evaluated the absolute density change in our samples as the changes observed are too small and the data does not allow us to reliably calculate absolute values.

## 4. Discussion

Our results provide detailed information on the density profile of ion tracks upon annealing and how temperature impacts ion track recovery and etching properties of the track in amorphous SiO$_2$. We have established the following facts:

1. The ion track damage in a-SiO$_2$ is healed during annealing. For 30 minutes annealing, the healing process sets in at temperatures as low as 250 ºC.
2. Chemical etching of ion tracks in a-SiO$_2$ leads to the formation of conical etch pits. With increasing annealing temperatures, the size of the etch pits decreases and the half-cone angle increases indicating a decreasing track etch rate.
3. The track etch rate as a function of depth is constant before annealing but decreases much quicker near the surface upon annealing, indicating quicker damage annealing closer to the surface. However, proximity to the interface of the SiO$_2$ layer and the Si substrate does not have an influence on track annealing.

4. A geometrical etching model using a variable track etch rate agrees very well with the experimentally observed results for the etch pit size and cone angle.

5. Annealing of ion tracks at different depths is described by different activation energies at the surface and deep inside the sample.

6. Oxidising atmosphere slightly promotes ion track repair up to a certain depth while a capping layer somewhat reduces it.

7. SAXS measurements of the radial track profile suggest that a small amount of material flows from the shell and the core region. The radial profile changes, but overall density does not change significantly across the entire track.

The annealing process in a-SiO$_2$ is significantly different compared to that in crystals such as quartz [70,71], apatite [12,13], and olivine [32,72]. In these materials, ion tracks are characterised by amorphous cylinders within a crystalline matrix. The radial density distribution profile of these tracks is defined by a relatively sharp transition zone from the amorphous track to the crystalline matrix. Upon annealing, the track radius decreases due to recrystallisation processes at the track-matrix interface. For the case of a-SiO$_2$, the density distribution profile is defined by an under-dense core and an over-dense shell region linked by a smooth transition region between them. Upon annealing of up to 600 °C, there is some flow of the material from the high-density region to the low-density region. The total radius of ion tracks defined by the core and shell regions almost remains the same. This flow behaviour is expected and consistent with the high viscosity of SiO$_2$ at 600 °C. We suspect that at higher temperatures, when the viscosity is sufficiently low, the flow will equalize the density difference. Despite the only small change in the track density profile, we observe a big change in the track etching rate already at much lower temperatures. We also find that track etching is significantly decreased towards the surface, demonstrating a significant role of the surface in the track recovery process.

To better understand the preferential annealing from the sample surface, we need to consider the etching mechanisms of pristine and track damaged a-SiO$_2$. While the etching mechanism of a-SiO$_2$ in aqueous hydrofluoric acid has not yet been fully resolved and only qualitative explanations of possible mechanisms have been provided, D. J. Monk *et al.* [73] and D. M. Knotter [74], pointed out that the Si-O bonds must be broken and replaced by Si-F bonds in order to etch amorphous silicon dioxide. This process can be accomplished through an electrophilic attack or a nucleophilic substitution process on the silicon-oxygen network. Furthermore, it was pointed out by L. Vlasukova *et al.* [75], and K. Awazu and H. Kawazoe [76], that the oxygen atoms in the Si-O network are displaced along the trajectory of swift heavy ions. Moreover, the irradiation creates small rings with decreased angle of Si-O-Si bonds. Thus, the Si-O bonds are stressed by the reconfiguration of the energetically favoured six-component SiO$_4$-tetrahedral ring into three- and four-component rings. The preferential etching of the ion track can be explained by the fact that strained Si-O bonds are energetically easier to break. The oxygen-deficient defects can be repaired during annealing in an oxygen atmosphere by the direct uptake of oxygen from the atmosphere which explains the influence of oxygen on the annealing process. Furthermore, we hypothesise that stress relaxation of the Si-O bonds is facilitated at the surface of the sample, in contrast to the interior. This stress relaxation, which is difficult at the SiO$_2$/Si interface, may therefore contribute to the quicker healing of ion tracks near the sample surface. We put forth the hypothesis that the observed track recovery process is significantly influenced by two primary factors: stress relaxation and oxygen diffusion. The interplay between these mechanisms can further elucidate the variation in activation energies (see section 3.3) at different initial depths.

## 5. Conclusions

In conclusion, SAXS measurements of ion tracks in $SiO_2$ provide accurate information about their radial density profile and about the complicated annealing dynamics. The transformation of ion tracks into conical etch pits by selective chemical etching reveals the complexity of the track profile with depth and the specific annealing process. Analysis of the size and shape of the etch pits as a function of temperature, annealing time and depth provides clear evidence that tracks are more readily annealed at the surface and are characterized by a smaller track etch rate within the first hundreds of nm from the surface. The fact that the annealing process is affected by the ambient environment as well as a significant difference observed for ion tracks annealed with and without capping layer suggests that the healing process is influenced by the presence or absence of oxygen. The activation energy for track annealing almost halves for deeper sample layers compared to the surface and explains the more effective healing process at the sample surface. Numerical simulations which consider a non-constant track etch rate provide a consistent description of the different etch pit shapes observed experimentally. The ion track not only anneals radially and with depth but also anneals at a faster rate from the surface. These results have immediate consequences for radioactive waste storage across the world. A study of track damage healing in borosilicate glasses based on these results and using methods explained here could have a direct impact on the rectification of damage in those glasses. This in turn could potentially allow for the storage of a higher weight percentage of nuclear waste, especially when the nuclear waste is stored on land and is easily accessible for annealing. The findings also suggest that the shape and structure of etch pits fabricated in silicon dioxide can be tuned depending on the application's requirements which provides new opportunities for the fabrication of solid-state nanopores in silicon dioxide membranes. The shape and size of conical nanopores influence the translocation properties of ions and molecules, such as ion/molecular selectivity, I-V characteristics, and electroosmotic flow rectification [77–80] and directly impact the enormous variety of applications.

## 6. Acknowledgments

The authors acknowledge the UNILAC staff at the GSI Helmholtz Center for Heavy Ion Research, Darmstadt (Germany) for carrying out the irradiations of thin films of a-SiO$_2$ on the UNILAC X0 beamline. Part of the research was undertaken at the SAXS/WAXS beamline at the Australian Synchrotron, part of ANSTO, and we thank the beamline scientists for their technical assistance. This work used the ACT node of the NCRIS-enabled Australian National Fabrication Facility (ANFF-ACT). This research was supported by an AINSE Ltd. Postgraduate Research Award (PGRA) and the Australian Government Research Training Program (RTP) Scholarship. The authors also acknowledge financial support from the Australian Research Council (ARC) under the ARC Discovery Project Scheme (DP180100068).

## 7. Data Availability

Data will be made available on request

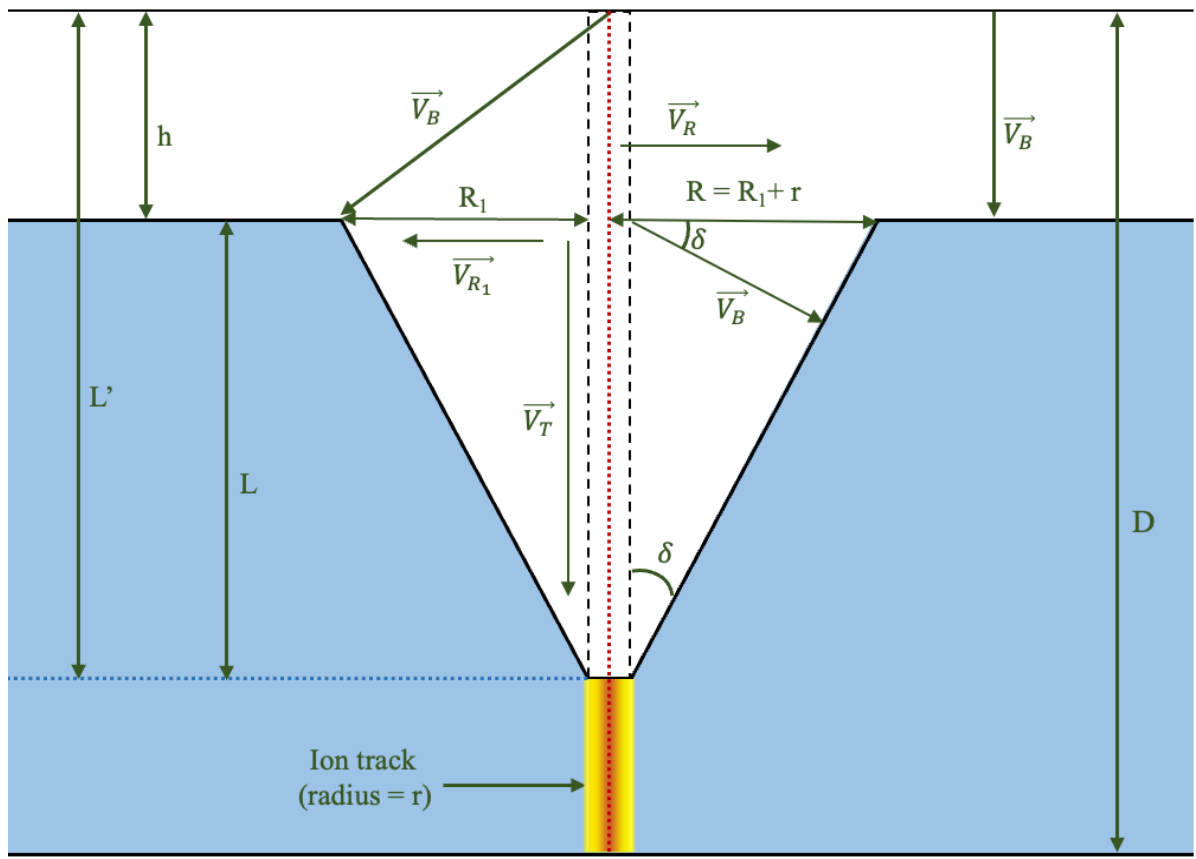

***Figure 1:*** *Schematics showing the formation process of the etch pit (by chemical track etching) of radius R from ion track of radius r. The shape and size of the etch pit depends on the track etch rate ($V_T$) and the bulk etch rate ($V_B$). The half cone angle of the etch pit is denoted by δ and the depth of the etch pit is represented by L. Bulk etching leads to removal of layer of thickness $h = L' - L$ from the top surface.*

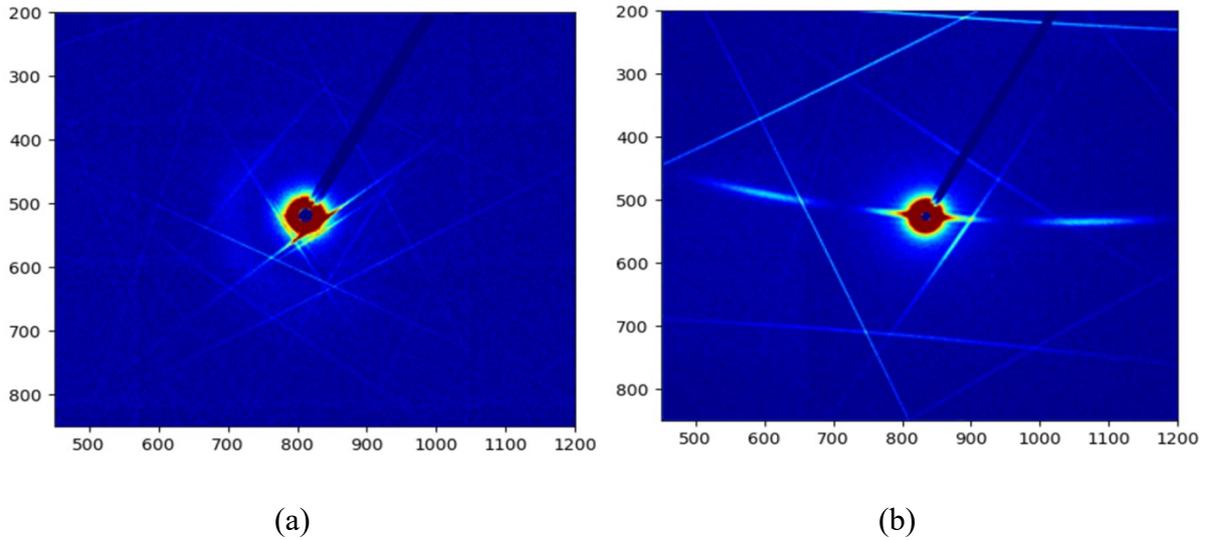

|(a)|(b)|

***Figure 2:*** *Two-dimensional scattering patterns (detector images) of parallel ion-tracks formed by 1.6 GeV $^{197}$Au ions in a-SiO$_2$: ion-track axis almost aligned (a) and titled by an angle of ~10° with respect to the X-ray beam (b). The scattering patterns correspond to a two-dimensional cut of the ion-tracks in reciprocal space. The central part of the detector is protected against the primary beam by a beam stop. The bright straight lines are ascribed to Kossel lines due to scattering from the crystalline Si substrate.*

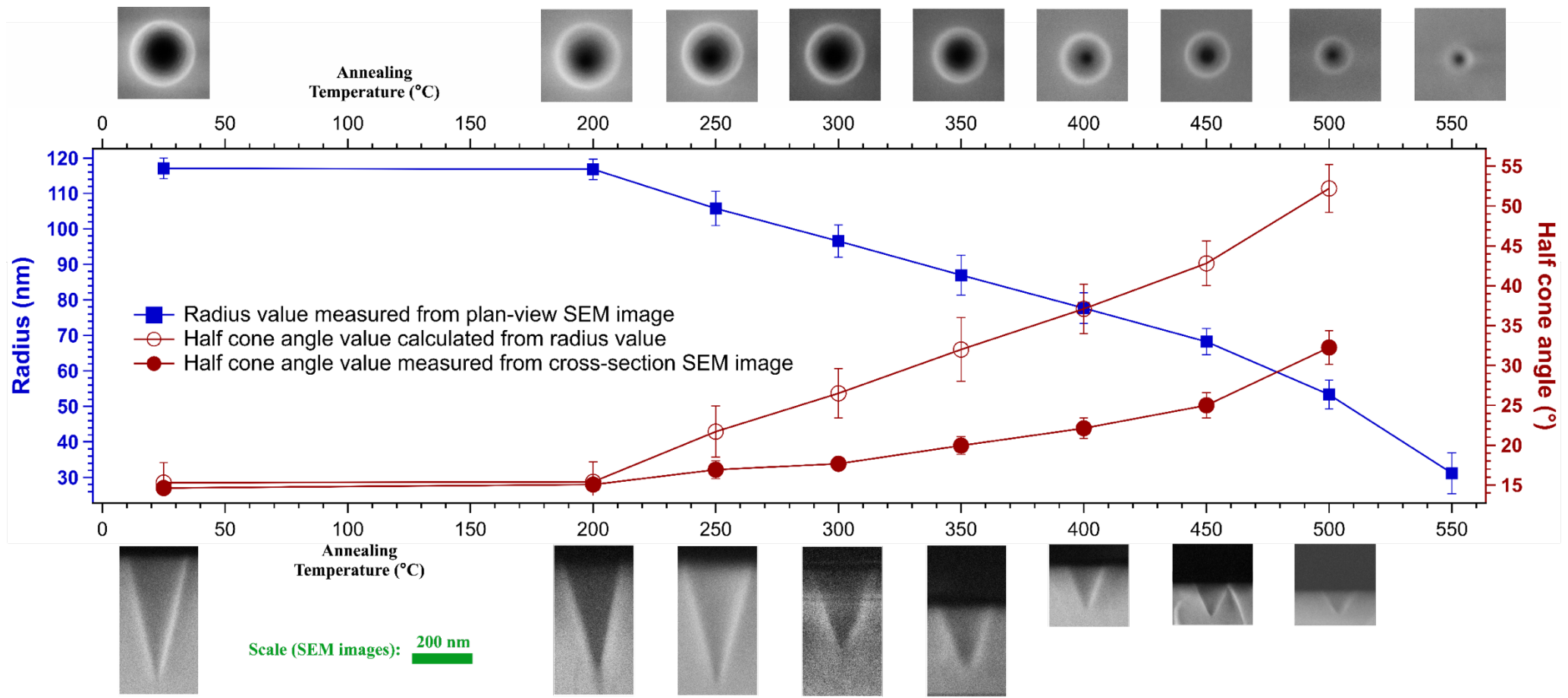

*Figure 3:* Radius (blue filled squares) and half cone angle (red filled circles) of the etch pits fabricated in a-SiO$_2$ deduced from SEM images as a function of the annealing temperature. The samples were annealed at different temperatures for 30 mins and subsequently etched for 10 mins in 3% HF. The etch pit radius and the length of the conical etch pits decrease, and the half cone angle increases with increasing annealing temperature. On top of the graph, plan-view SEM images of the etch pits and at the bottom, cross-section SEM images of the etch pits are shown corresponding to their annealing temperature. The scale bar applies to all SEM images. Half cone angle values (red open circles) calculated from the surface radius values employing the track-etching model as a function of annealing temperature are also shown. The solid lines are guides to the eye.

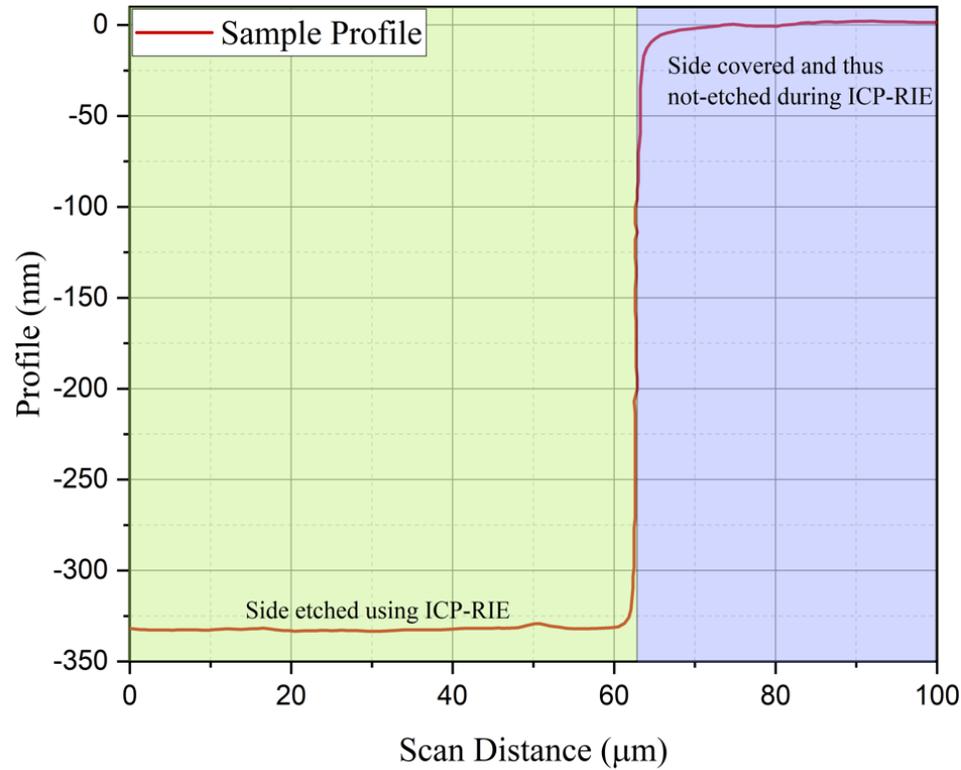 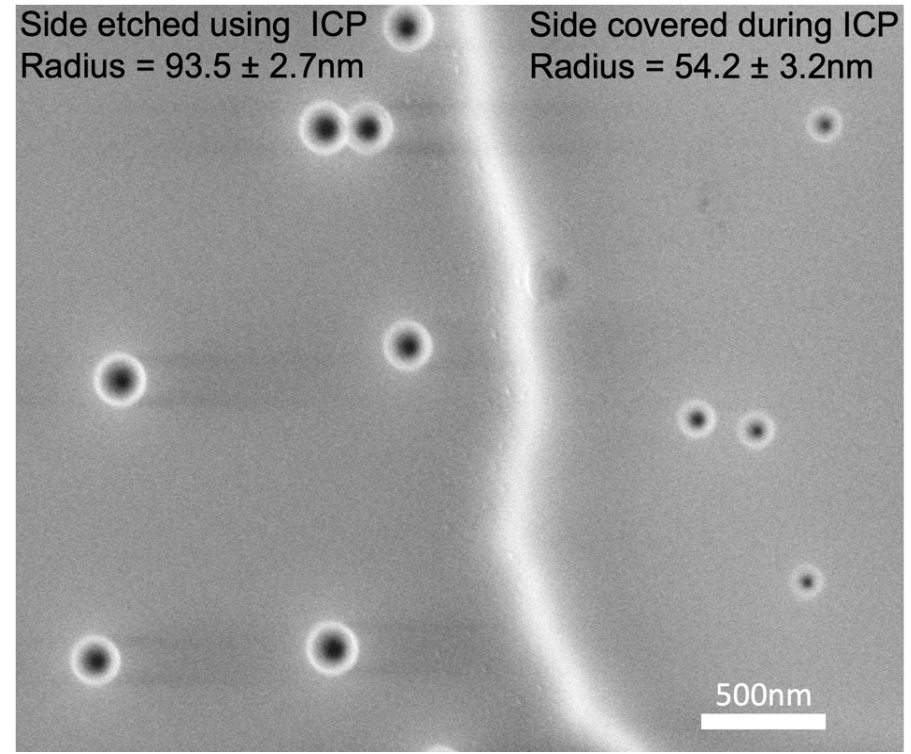

(a) (b)

*Figure 4:* (a) One-dimensional surface profile of a-SiO$_2$ sample (irradiated with 1.6 GeV Au ions with fluence of $1 \times 10^8$ ions cm$^{-2}$) annealed at 500 °C for 30 min. The right part of the sample was covered with polyimide film to protect it during ICP-RIE treatment which removed ~330 nm of the unprotected layer on the left. (b) Plan-view SEM image showing the etch pits in the ICP-RIE processed sample fabricated by etching the sample in 3% HF solution for 10 mins. On the left side (ICP-RIE processed side), the etch pits are 72.5% larger than on the protected right side of the sample.

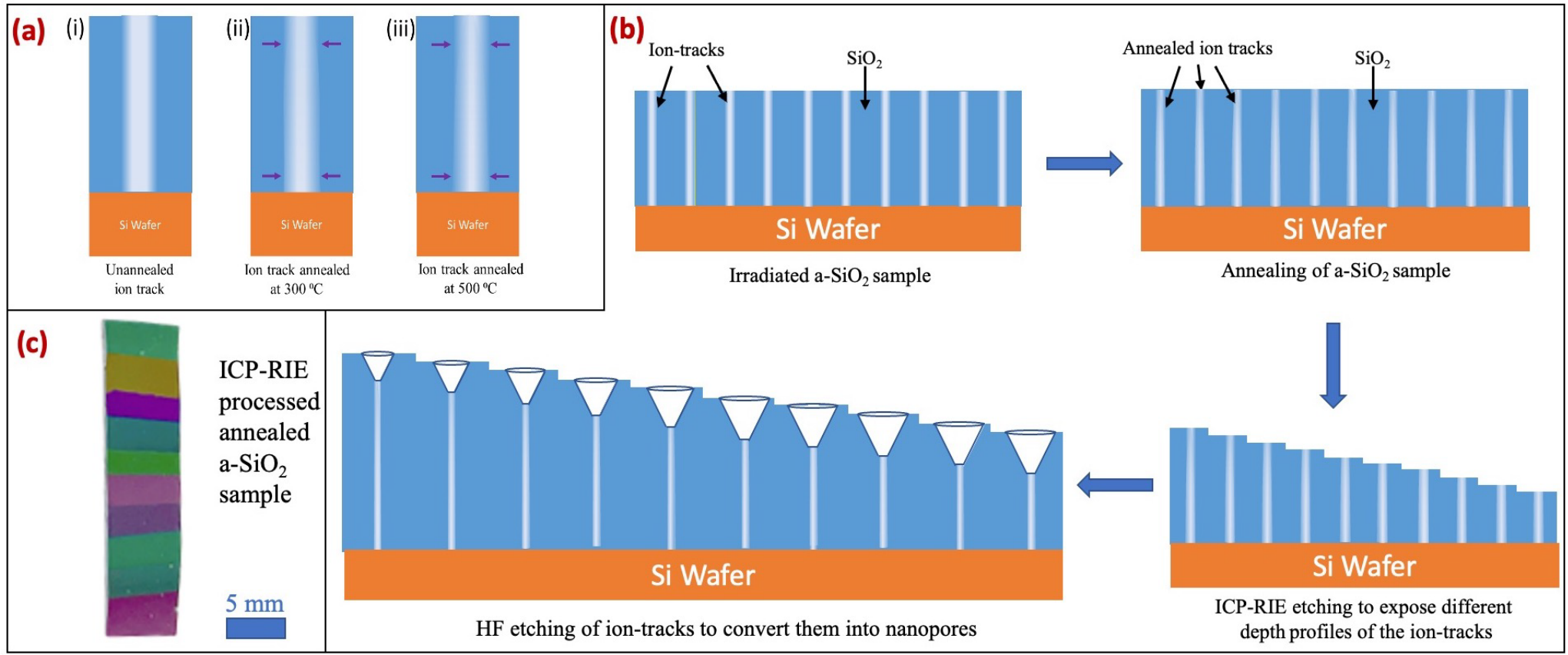

*Figure 5:* (a) Illustration of changing ion track profile radially as well as with depth. Profile of unannealed ion tracks (i), ion tracks annealed at 300 °C (ii) and ion tracks annealed at 500°C (iii) is shown. As indicated, the ion track anneals faster from the surface. (b) Illustration showing the processing of sample for the study of changing ion track profile with depth. The samples are first annealed and then 9 steps are created using ICP-RIE to expose different parts of the changing ion track. Subsequently, the ion tracks were etched to fabricate etch pits which were analysed using SEM. (c) Photo of an ICP-RIE processed sample. Ten different layer colours correspond to different thickness of the a-SiO$_2$ thin film. The illustrations shown are not to scale and designed for the understanding of the reader.

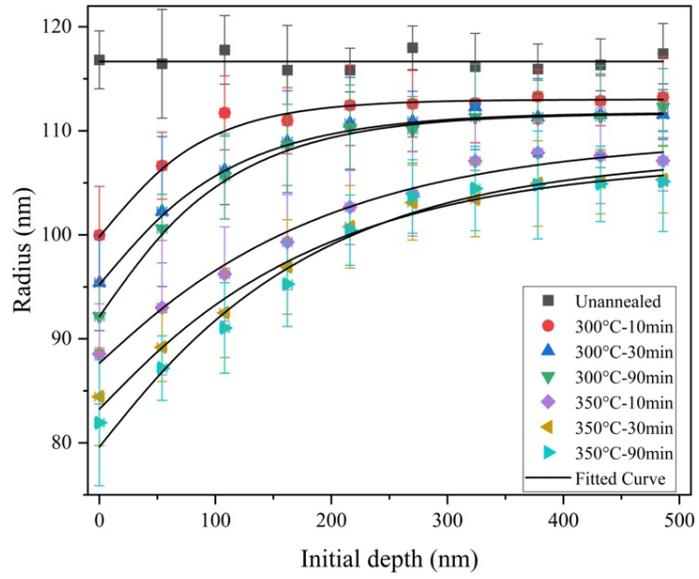 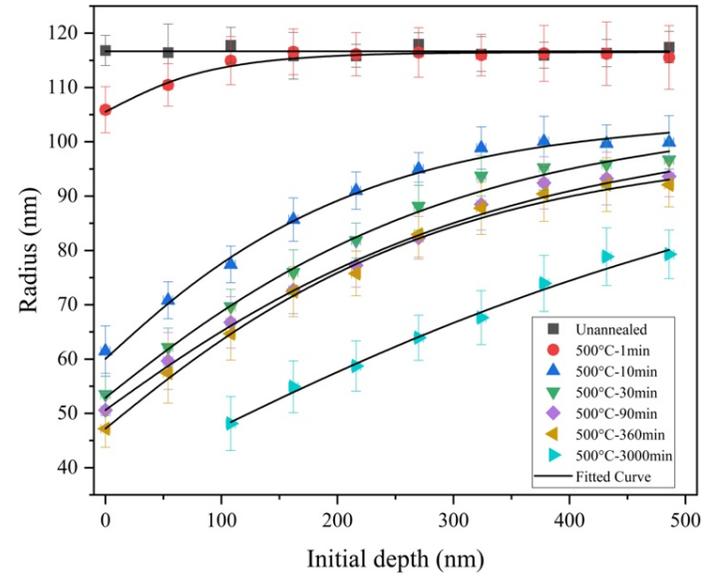

(a) (b)

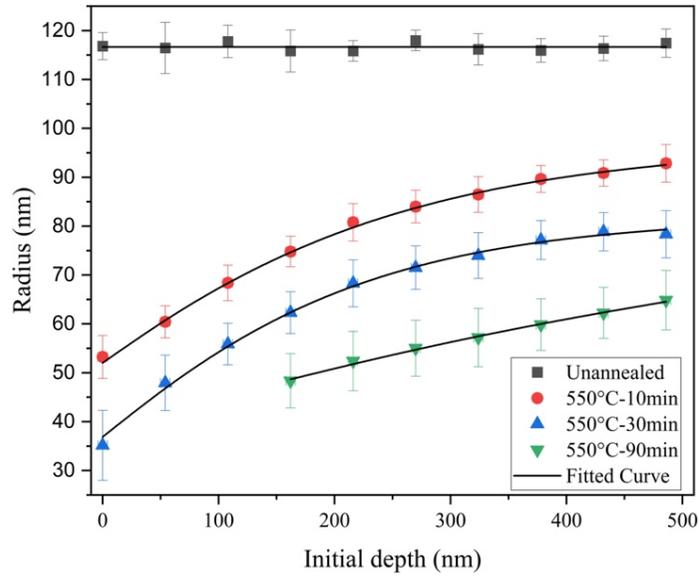 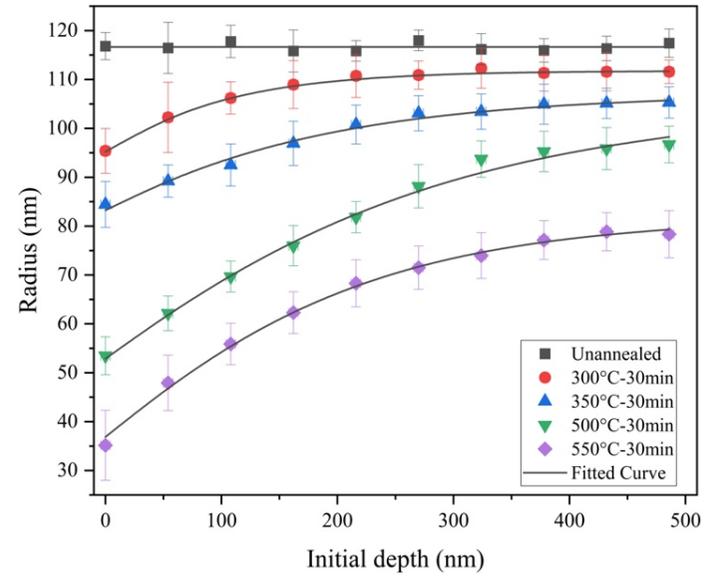

(c) (d)

*Figure 6:* Mean radius of the etch pits (deduced from SEM images) as a function of depth from initial surface for different annealing temperatures and times: samples annealed at 300 °C and 350 °C (a), at 500 °C (b) and 550 °C (c) . Comparison of different annealing temperatures for same annealing time (30 min) is shown in (d). Radius of unannealed tracks is shown in each plot for easy comparison. For all annealing conditions, the radius increases with increasing initial depth indicating a change in the damage profile of the ion track. Fits to the data obtained using Gompertz function are shown through solid black lines. The fit parameters are given in table S1.

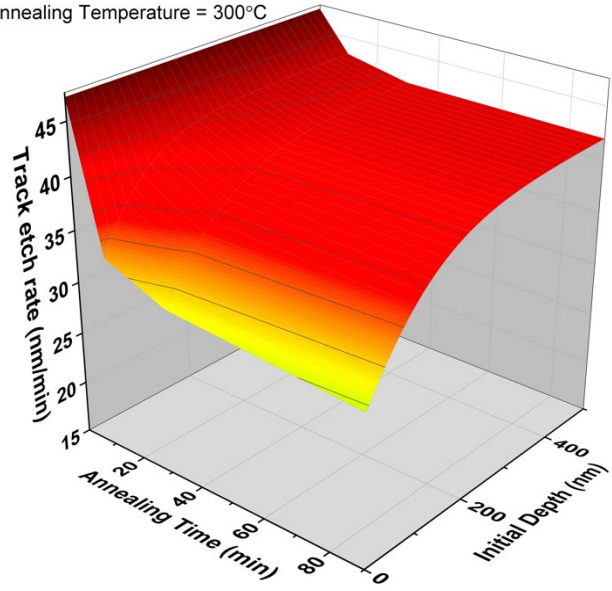
(a)

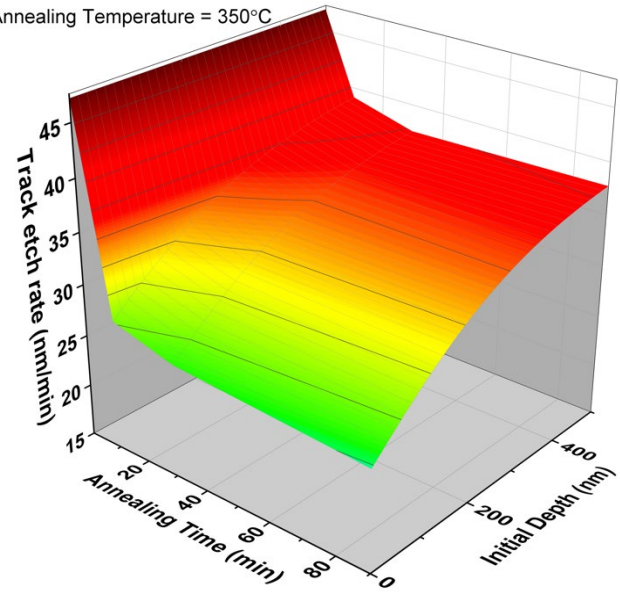
(b)

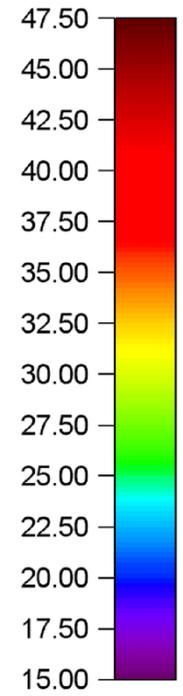

Track etch rate (nm/min)

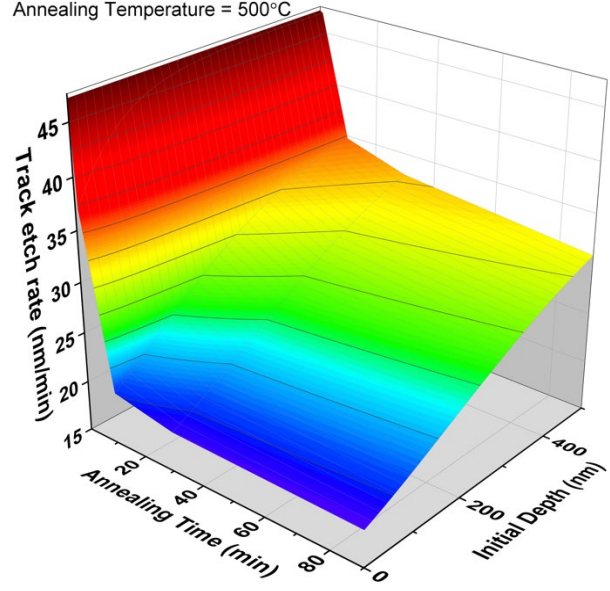
(c)

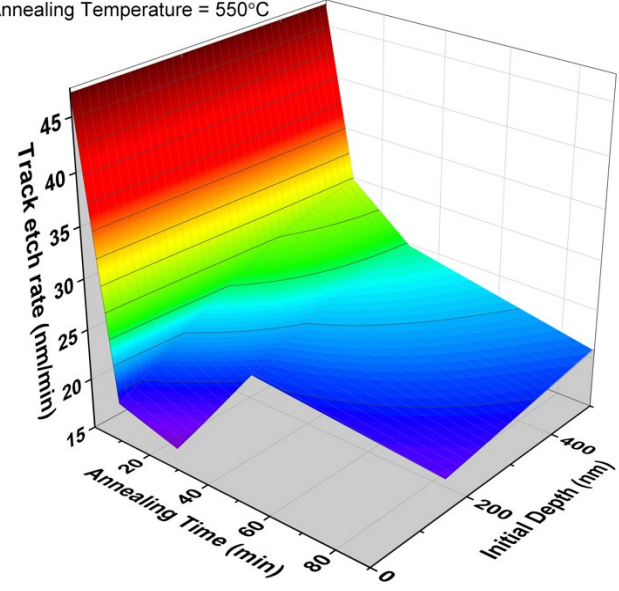
(d)

*Figure 7:* Three-dimensional plots showing the track etch rate as a function of annealing time and initial depth for samples annealed at a temperature of (a) 300 °C, (b) 350 °C, (c) 500 °C, and (d) 550 °C. The change in the track etch rate decreases with increasing annealing time. The track etch rate increases with increasing sample depth for any annealing duration, and appears to level beyond a certain depth, which is dependent on annealing time and temperature.

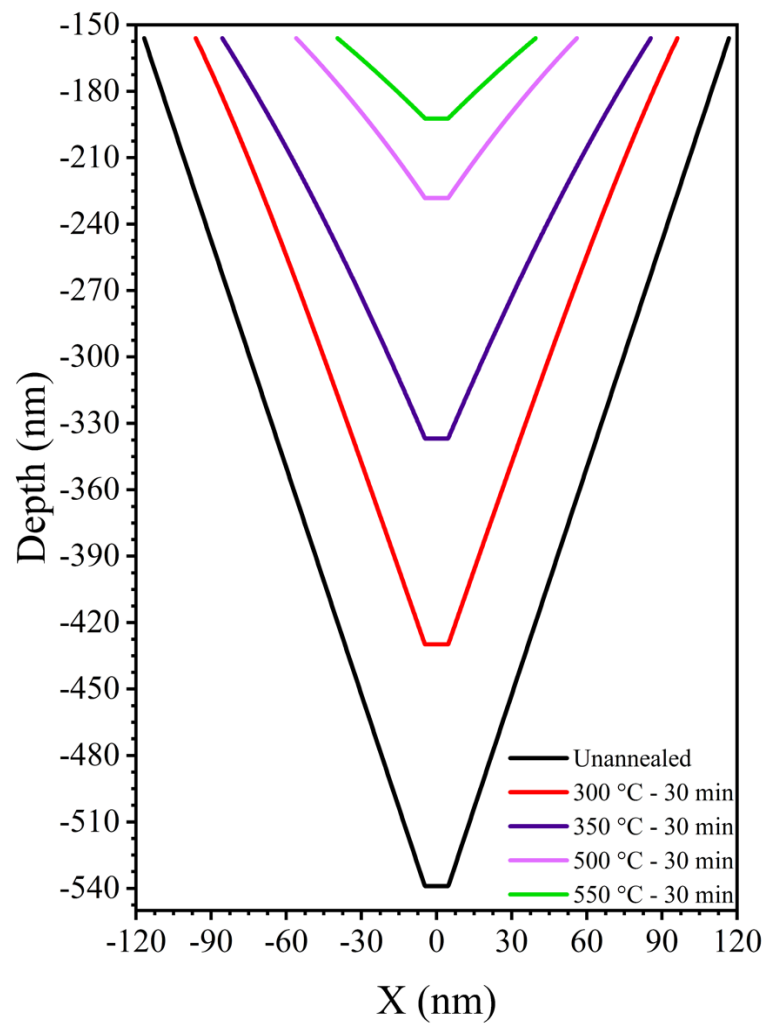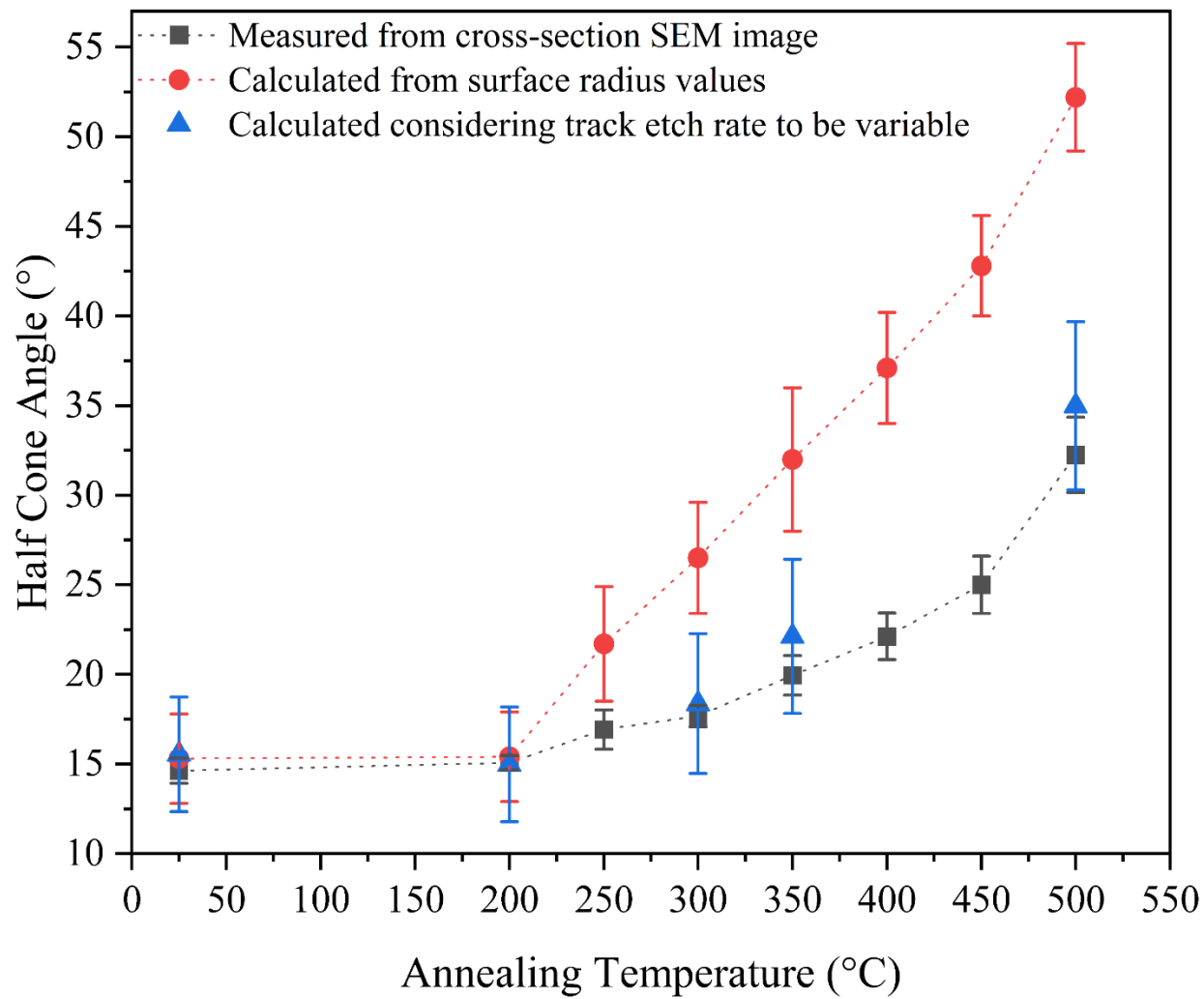

(a)                      (b)

*Figure 8: (a) Simulation of the etch pit shape for different annealing conditions considering a variable track etch rate and an etching time of 10 min in 3% HF. The etch pits start at -156 nm because of the removal of the top layer by bulk etching. (b) Half cone angle values (measured from cross-section SEM images, calculated from the surface radius values (as shown in figure 3) of the etch-pits and calculated considering the track etch rate to be variable) as a function of annealing temperature. The half cone angle values calculated using the surface radius values deviate significantly from the measured values. On*

*the other hand, within the margin of error, the estimated half cone angle values and the shape and size depicted in (a) accord well with experimentally measured values.*

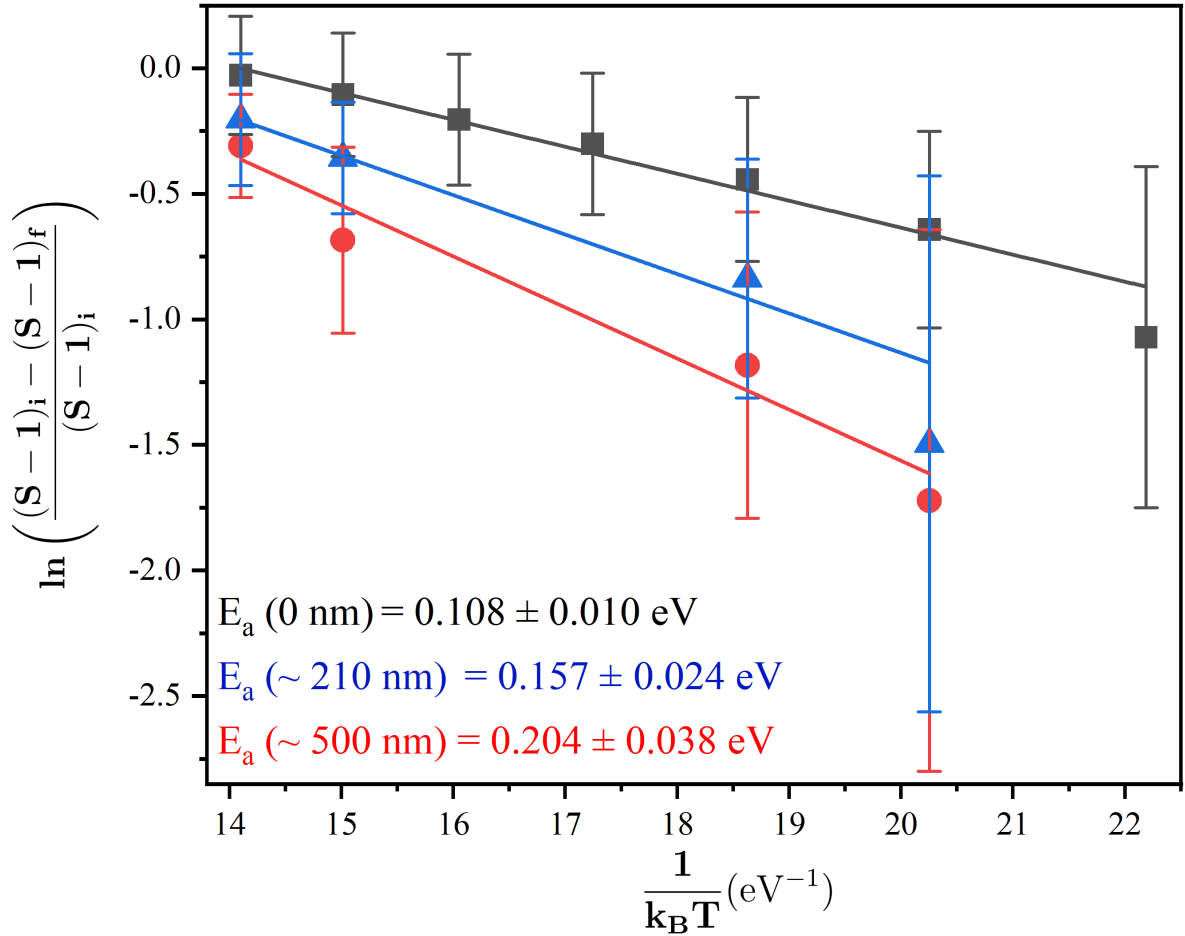

***Figure 9:*** *Variation of $\ln\left(\frac{(S-1)_i - (S-1)_f}{(S-1)_i}\right)$ vs. $\frac{1}{k_B T}$ for annealing of ion tracks in a-SiO$_2$ for the case of initial depths of 0 nm, ~210 nm, and ~500 nm, respectively. The solid lines are linear fits to the data. The activation energy $E_a$ is deduced from the slope of the linear fit and decreases with increasing initial depth.*

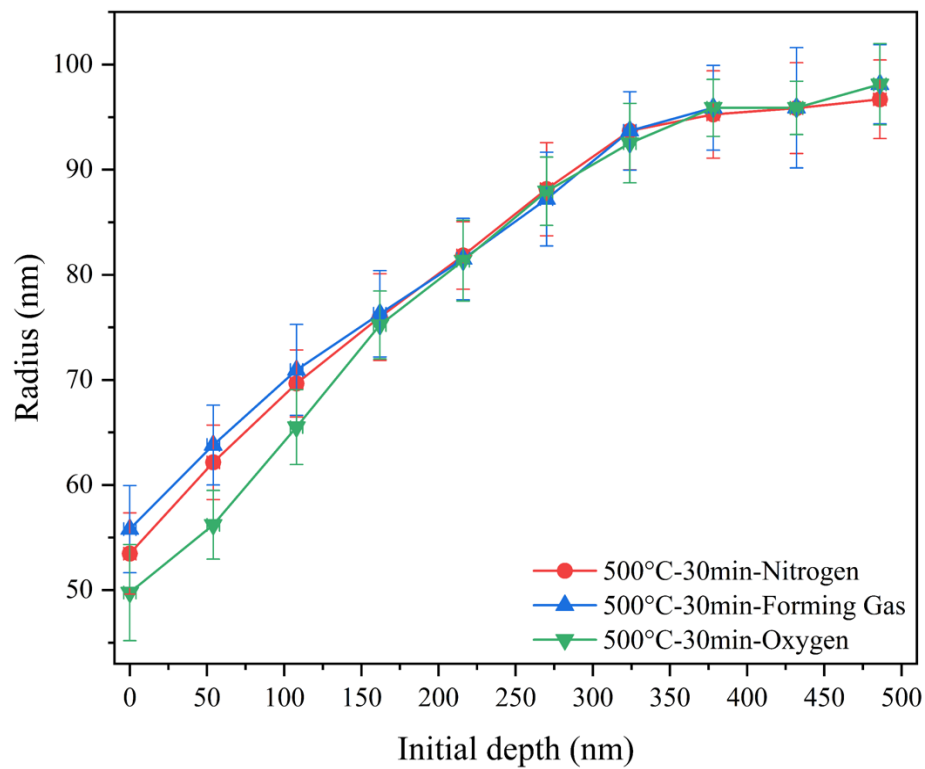
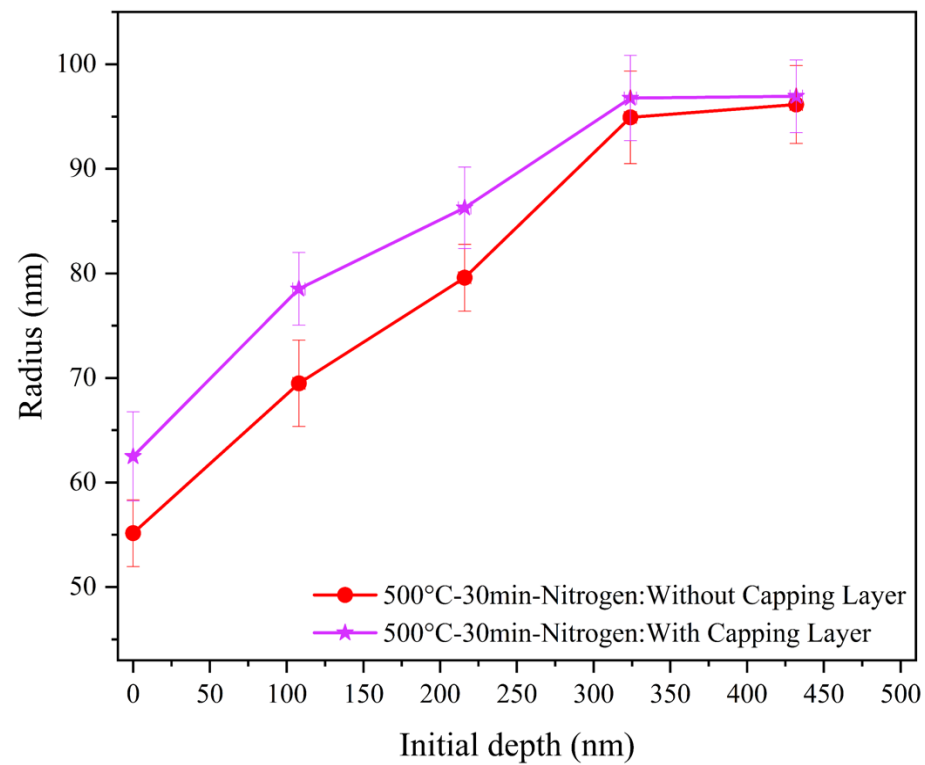

(a)            (b)

*Figure 10: (a) Radius of the etch pits as a function of initial depth for annealing at 500 ºC for 30 mins in different ambient atmospheres (dry nitrogen gas, dry oxygen and dry forming gas). (b) Radius of the etch pits for samples annealed at 500 ºC for 30 min in a dry nitrogen atmosphere with and without a 150 nm thick aluminium capping layer as a function of initial depth. The solid lines are a guide to the eye.*

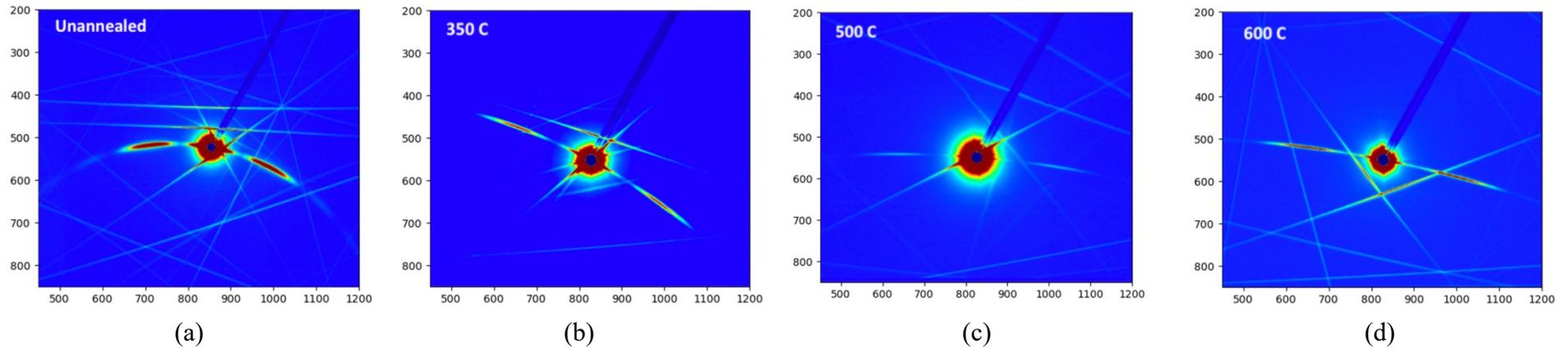

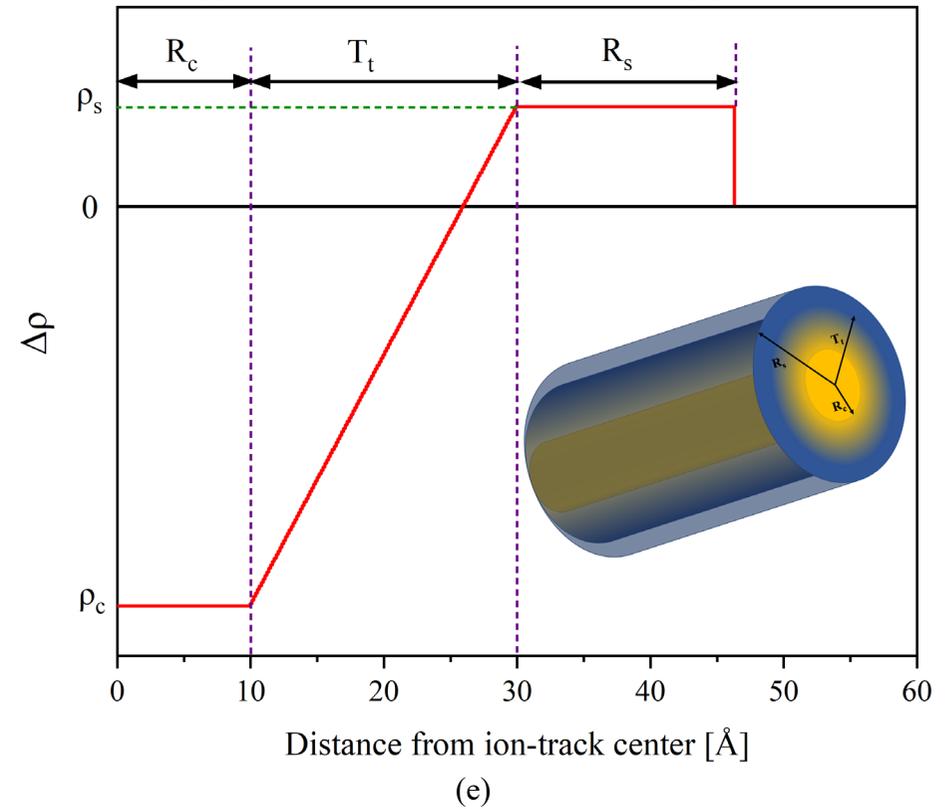

*Figure 11:* Detector images (2D scattering patterns) of ion tracks in amorphous silicon dioxide thin films: unannealed (a), annealed at 350 °C (b), at 500 °C (c) and at 600 °C (d) for a duration of 30 mins. Samples were titled by an angle of 5 or 10 degrees with respect to the incoming X-ray beam. (e) Profile of radial electron density distribution profile of the Core Transition Shell model. The total radius of the ion track is given by the sum of the core region ($R_c$), transition region ($T_t$), and shell region ($R_s$); $R = R_c + T_t + R_s$.

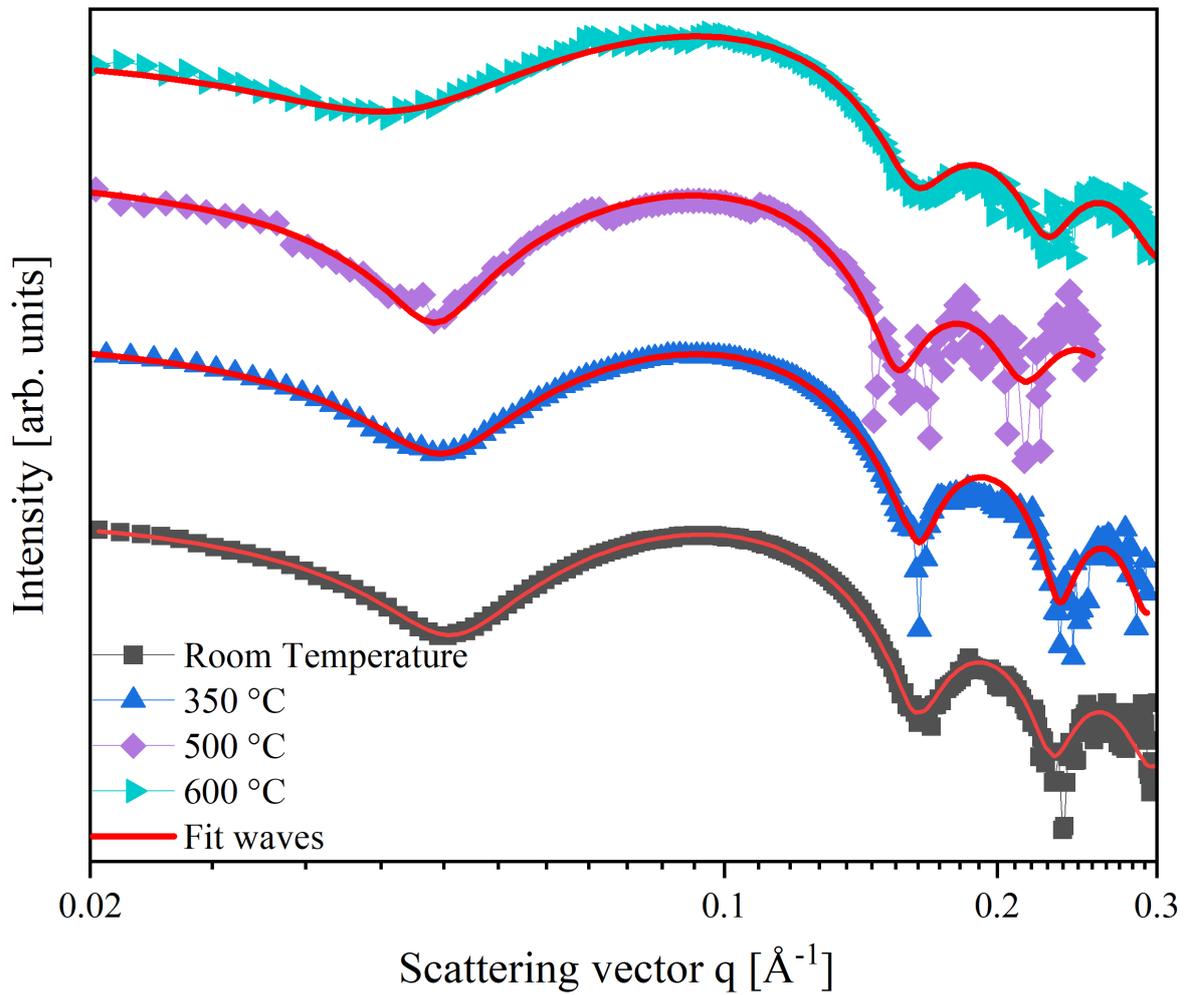

*Figure 12: SAXS scattering intensities from unannealed and annealed ion tracks in a-SiO$_2$. The samples were irradiated with 1.6 GeV Au ions with a fluence of $5 \times 10^{11}$ ions cm$^{-2}$ and annealed at various temperatures for a duration of 30 min. The SAXS intensity patterns are offset vertically for easy visualization. The solid red lines represent numerical fits obtained using the Core Transition Shell model.*

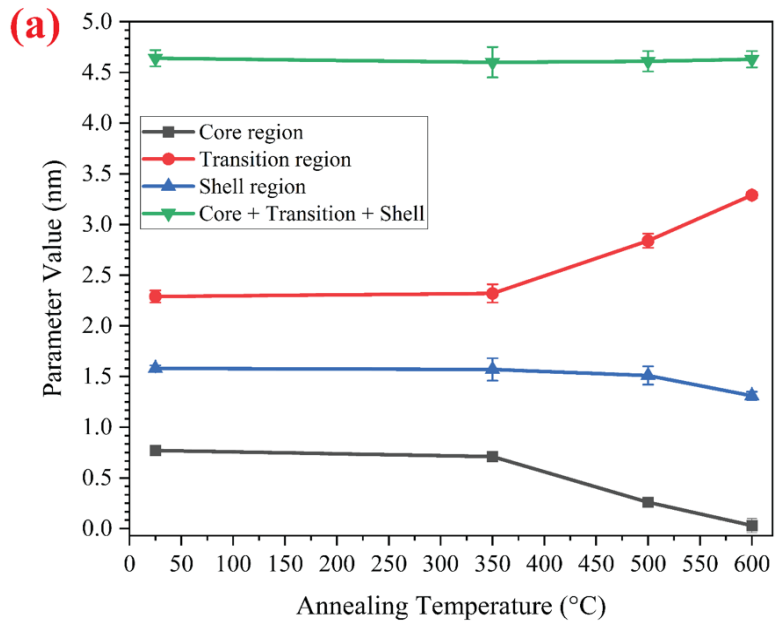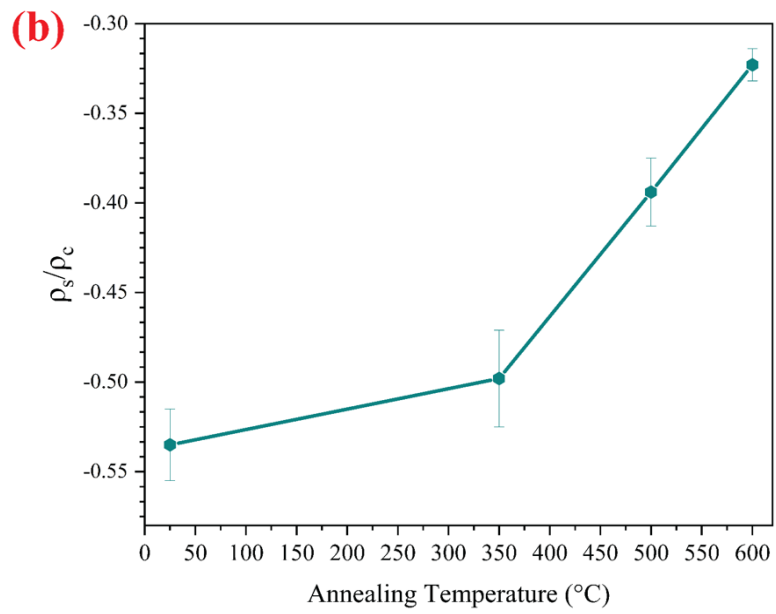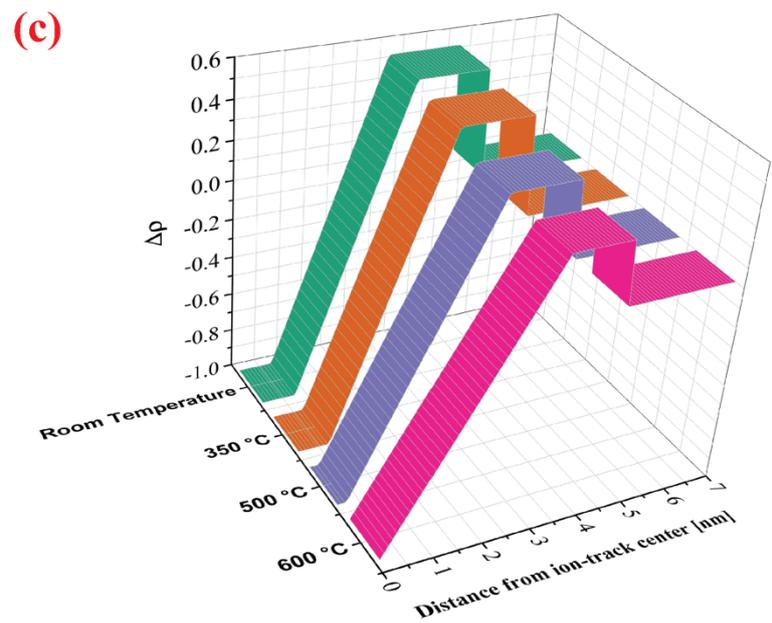

*Figure 13:* (a) Fitting parameters (core region ($R_c$), transition region ($T_t$), shell region ($R_s$), and radius of ion tracks ($R_c + T_t + R_s$)) from SAXS measurements for an unannealed a-SiO$_2$ sample and samples annealed at 350 °C, 500 °C, and 600 °C. (b) The ratio of the density of core region ($\rho_c$) to the density of shell region ($\rho_s$) as a function of annealing temperature. (c) The radial density distribution for different unannealed and annealed a-SiO$_2$ samples.

*Table 1:* Fitting parameters from SAXS measurements of unannealed and annealed ion tracks in a-SiO$_2$: core region ($R_c$), transition region ($T_t$), shell region ($R_s$), dispersity ($\sigma$), and radius of ion tracks ($R_c + T_t + R_s$). The ratio of the density of the core region ($\rho_c$) to the density of the shell region ($\rho_s$) has also been given.

| Annealing temperature (°C) | Core region $R_c$ (nm) | Transition region $T_t$ (nm) | Shell region $R_s$ (nm) | Dispersity $\sigma$ (nm) | $\rho_s/\rho_c$ | Radius $R_c + T_t + R_s$ (nm) |
|---|---|---|---|---|---|---|
| Unannealed | 0.77 ± 0.05 | 2.29 ± 0.06 | 1.58 ± 0.03 | 0.07 ± 0.01 | −0.54 ± 0.02 | 4.64 ± 0.08 |
| 350 | 0.71 ± 0.05 | 2.32 ± 0.09 | 1.57 ± 0.11 | 0.08 ± 0.01 | −0.50 ± 0.03 | 4.60 ± 0.15 |
| 500 | 0.26 ± 0.03 | 2.84 ± 0.07 | 1.51 ± 0.09 | 0.10 ± 0.01 | −0.39 ± 0.01 | 4.61 ± 0.10 |
| 600 | 0.04 ± 0.03 | 3.28 ± 0.03 | 1.31 ± 0.04 | 0.14 ± 0.02 | −0.32 ± 0.01 | 4.63 ± 0.08 |